\documentclass[fleqn,usenatbib,useAMS]{mnras}
\usepackage{newtxtext,newtxmath}

\usepackage[T1]{fontenc}

\DeclareRobustCommand{\VAN}[3]{#2}
\let\VANthebibliography\thebibliography
\def\thebibliography{\DeclareRobustCommand{\VAN}[3]{##3}\VANthebibliography}

\numberwithin{equation}{section}

\usepackage{appendix}
\usepackage{lscape}
\usepackage{ulem}
\usepackage[detect-none]{siunitx}
\usepackage{tabularx}
\usepackage{graphicx}	
\usepackage{amsmath}	
\usepackage{natbib}
\usepackage{lineno}
\usepackage[usestackEOL]{stackengine}

\newcommand{\microJybm}{\,\mathrm{\umu Jy\,beam^{-1}}}
\newcommand{\microJy}{\,\mathrm{\umu Jy}}

\newcommand{\x}{{$\times\,$}}
\newcommand{\plus}{$\pm$}
\newcommand{\emerlin}{\textit{e}{-MERLIN}}
\newcommand{\flux}{$\,\mathrm{\umu Jy}$}

\title[SPARCS-North Wide-field VLBI Survey]{SPARCS-North Wide-field VLBI Survey: Exploring the resolved $\mu$Jy extra-galactic radio source population with EVN+e-MERLIN}

\author[Njeri~A. et al.]{Ann~Njeri,$^{1,2}$\thanks{E-mail: ann.ngendo@postgrad.manchester.ac.uk}
Robert~J.~Beswick,$^{1}$
Jack~F.~Radcliffe,$^{1,3,4}$
A.~P.~Thomson,$^{1}$
N.~Wrigley,$^{5}$ T.~W.~B.~Muxlow,$^{1}$ 
\newauthor
M.~A.~Garrett,$^{1}$
Roger.~P.~Deane,$^{2,6}$ Javier~Moldon,$^{7}$ Ray~P.~Norris,$^{8,9}$ and Roland~Kothes$^{10}$\\
$^{1}$Jodrell Bank Centre for Astrophysics, University of Manchester, Oxford Road, Manchester M13 9PL, UK\\
$^{2}$School of Mathematics, Statistics and Physics, Newcastle University, Newcastle upon Tyne NE1 7RU, UK\\
$^{3}$Department of Physics, University of Pretoria, Lynnwood Road, Hatfield, Pretoria 0083, South Africa\\
$^{4}$National Institute for Theoretical and Computational Sciences (NITheCS), 1 Jan Smuts Ave, Braamfontein, Johannesburg, 2000, South Africa\\
$^{5}$Jodrell Bank Observatory, Lower Withington, Bomish Ln, Macclesfield SK11 9DW, UK\\
$^{6}$Wits Centre for Astrophysics, School of Physics, University of the Witwatersrand, Private Bag 3, 2050, Johannesburg, South Africa\\
$^{7}$Instituto de Astrof\'isica de Andaluc\'ia (IAA, CSIC), Glorieta de las Astronom\'ia, s/n, E-18008 Granada, Spain \\
$^{8}$Western Sydney University, Locked Bag 1797, Penrith South DC, NSW 2751, Australia \\
$^{9}$CSIRO Space and Astronomy, Australia Telescope National Facility, PO Box 76, Epping, NSW 1710, Australia \\
$^{10}$Dominion Radio Astrophysical Observatory, 1 Observatory Crescent, Ottawa, ON K1A 0C6, Canada
}

\date{Accepted 2022 November 27. Received 2022 November 27; in original form 2022 August 04}

\pubyear{2022}

\begin{document}
\label{firstpage}
\pagerange{\pageref{firstpage}--\pageref{lastpage}}
\maketitle

\begin{abstract}
The SKA PAthfinder Radio Continuum Surveys (SPARCS) are providing deep-field imaging of the faint (sub-mJy) extra-galactic radio source populations through a series of reference surveys. One of the key science goals for SPARCS is to characterize the relative contribution of radio emission associated with AGN from star-formation (SF) in these faint radio source populations, using a combination of high sensitivity and high angular resolution imaging over a range of spatial scales (arcsec to mas).  To isolate AGN contribution from SF, we hypothesise that there exists a brightness temperature cut-off point separating pure AGN from SF. We present a multi-resolution (10–100\,mas) view of the transition between compact AGN and diffuse SF through a deep wide-field EVN+\emerlin{}, multiple phase centre survey of the centre of the Northern SPARCS (SLOAN) reference field at 1.6\,GHz. This is the first (and only) VLBI (+\emerlin{}) milliarcsecond angular resolution observation of this field, and of the wider SPARCS reference field programme. Using these high spatial resolution ($9$\,pc -- $0.3$\,kpc at $z\sim1.25$) data, 11 milliarcsecond-scale sources are detected from a targeted sample of 52 known radio sources from previous observations with the \emerlin{}, giving a VLBI detection fraction of $\sim 21\%$. At spatial scales of $\sim9\,$pc, these sources show little to no jet structure whilst at $\sim 0.3\,$kpc one-sided and two-sided radio jets begin to emerge on the same sources, indicating a possible transition from pure AGN emissions to AGN and star-formation systems.


\end{abstract}




\section{Introduction}
The SKA PAthfinder Radio Continuum Surveys (SPARCS\footnote{\url{http://spacs.pbworks.com}}) are providing deep-field imaging with multiple SKA pathfinders. These surveys draw upon the unique and complementary qualities of various SKA pathfinder/precursor facilities operating at various resolutions and frequencies, while testing various technical aspects between different instruments \citep{Norris2013,Norris2017,Simpson2017}.  Using next-generation and upgraded radio telescopes such as the LOFAR, ASKAP, MeerKAT, APERTIF, EVLA and \emerlin{}, the SPARCS programme is undertaking a series of deep wide-field reference radio continuum surveys to study the formation and evolution of galaxies, cosmological parameters and the large-scale structures driving them. To utilize these observations, different teams across the globe are developing different techniques such as multi-scale deconvolution, source extraction, identification and classification, and multi-wavelength cross-identification \citep[e.g.,][]{Kondapally2021,Inigo2022}. These observations, making use of current radio facilities have confirmed that the extra-galactic SKA-sky will be dominated by a mixture of intense star forming galaxies (SFGs), radio-quiet and radio-loud AGN \citep[e.g.,][]{Morabito2017,Macfarlane2021,An2021,Whittam2022}. These sources are typically situated between redshifts of 0.5 and 4 with sub-mJy flux densities and in many cases, are luminous merging galaxies often hosting co-evolving SF and accretion activity \citep[e.g.,][]{Muxlow2020}. Characterising this radio population with existing observations is vital to guide development of key SKA science programmes, a key scientific objective of the SPARCS programme.

At lower flux densities $S_\mathrm{1.4GHz}\leq 100 \mu$Jy,
SFGs tend to dominate over faint AGN. Consequently, there is a requirement to disentangle and distinguish between emissions from AGN accretion and star-formation \citep[e.g.,][]{Padovani2016,Vernstrom2016,Vlugt2021}. Whilst  multi-wavelength astronomical approaches such as FIRRC \citep[e.g.,][]{Thomson2014}, IRAC colours \citep[e.g.,][]{Rawlings2015} and optical emission line diagnostics \citep[e.g.,][]{Law2021} are making great strides in attempting to characterise these processes, they are not available for every source \citep[e.g.,][]{Algera2020}. Furthermore, many of the current radio surveys, including the present SKA pathfinder surveys, are still limited by their angular resolution and are unable to morphologically distinguish all AGN from SFGs. Therefore, a robust method of distinguishing between accretion and star-formation is essential, independent of the biases associated with multi-wavelength classification \citep[e.g.,][]{Radcliffe2021}. 

High spatial resolution wide-field radio observations provide a key diagnostic for distinguishing between accretion and star-formation and importantly, spatially resolve and quantify their contributions within individual galaxies \citep[e.g.,][]{Morabito2022}. The Very Long Baseline  Interferometry (VLBI) distinctly provides high spatial resolution scales, and offers a combination of high point source sensitivity and low surface brightness sensitivity which restricts the selection effect to small compact emissions. For example, at $z = 1$, $1\arcsec$ corresponds to $\sim8\,$kpc, while $10\,$mas provided by VLBI corresponds to $\sim8\,$pc. This unique VLBI characteristic is critical in identifying high brightness temperature ($T_B > 10^5\,$K) emissions from compact radio cores and thus constraining emissions from AGN in these extragalactic radio source populations \citep[e.g.,][]{Middelberg2013,Herrera2017,Herrera2018,Radcliffe2018,Radcliffe2021}.

\section{VLBI detection rate vs resolution}
We consider VLBI a clear cut indicator for the presence of accretion activity \citep{Condon1991,Middelberg2008}, with current VLBI studies of the high redshift ($z\sim2$) radio source population having made tremendous strides in isolating AGN contribution from star formation at $\sim$ parsec scales \citep[e.g.,][]{Chi2013,Herrera2017,Herrera2018,Radcliffe2018,Radcliffe2021}. However, a gap remains whereby we know little of the radio morphology of these objects at sub-kpc scales and thus the transition point between accretion emission processes associated with compact cores and diffuse/extended emission processes associated with star formation remains unclear \citep[e.g.,][]{Rees2016,Hardcastle2020}. To characterize the true nature of these sources and to isolate AGN contribution from star formation, we require intermediate resolution imaging at sub-kpc scales. Since no non-AGN process (except for the very rare radio supernovae) can exceed $T_B > 10^5\,$K, we hypothesise that there exists a brightness temperature cut-off point separating pure AGN from SF. To test this hypothesis, we need to achieve and test various angular resolutions between 10--200\,mas corresponding to various intermediate spatial (sub-kpc) scales at which a radio inteferometric instrument (+VLBI) becomes sensitive to compact radio emission purely attributed to AGN activity in the host galaxy. This requires optimizing angular resolutions of these VLBI surveys at intermediate spatial scales, with sufficient sensitivity to capture all AGN contribution and to resolve out all diffuse emissions. This is important if we are to; i) trace the pure spatial extent of star formation through cosmic time, and, ii) understand the radio emission mechanisms- related to accretion processes in the central core engines which drive the feedback mechanisms in the host galaxy. We also note while a detection at a sufficiently high resolution certainly implies an AGN, the converse is not always true: even a radio-loud AGN does not necessarily yield a VLBI detection, and this is not a function of source strength or sensitivity \citep{Rees2016}.

Combining \emerlin{}+VLBI provides a unique prospect of imaging at these intermediate spatial scales. Particularly, combining the EVN+eMERLIN will allow characterization of the $\microJy$ radio regime from milliarcsecond scales to $0\farcs2$ ($\sim 10 - 200\,\mathrm{mas}$). The \emerlin{} antennas increase the \textit{uv}-coverage at shorter baselines, thus characterising the source/jet structure at sub-kpc scales while the EVN longer baselines recover the compact cores at parsec scales. On the other hand, the VLA provides the largest spatial (>kpc) scales at $\mathrm{1-2\,GHz}$ required for total flux density measurements. Therefore, a combination of these observations with VLBI will provide kpc--parsec imaging spatial scales, essential in probing the most compact radio sources and their accompanying radio jets/structures. This enhanced imaging fidelity will enable the detection of low luminosity radio sources and provide an insight into their morphology. The derived radio spectrum will be useful, for example, in determining whether they are associated with broad absorption lines or thermal radiation for a better understanding of the nature of these radio-quiet objects \citep[e.g.,][]{Blundell2007}.

Consequently, we present a wide-field VLBI survey of the SPARCS-North field, by combining (for the first time) the EVN+eMERLIN spatial scales ranging from $\sim$ 10--100\,mas. This is one of the first multi-resolution studies of the faint radio source population at 10--100\,mas. Using the multi-resolution data and the intermediate spatial scales obtained in this work, we study the transition from compact radio emissions purely associated with AGN to diffuse radio emissions associated with either AGN with jets, star-formation or both jets and star-formation. We detected 11 VLBI sources at $1\sigma$ sensitivity of $\sim 6\,\microJybm$ with $\sim14\,$h on target. This paper is outlined as follows: we describe the survey design and data reduction strategies in Section \ref{s:datareduction}, including the primary beam corrections for the EVN+\emerlin{} array in Section \ref{PBC}, and the additional complementary data from the deep wide-field \emerlin{} and VLASS radio surveys. In Section \ref{s:analysis}, we present analysis of radio properties at 10-100 mas multi-resolution coverage, which corresponds to 9\,pc--0.28\,kpc (assuming $z\sim1.25$) spatial scales. We give a summary of this work and outline the plans for a future further wide-field VLBI survey to increase the source sample and sensitivity in Section \ref{s:conclusions}. Throughout this paper, we adopt the $\Lambda$CDM cosmology parameters with $H\mathrm{_0} = 67.4$\,kms$^{-1}$\, Mpc$^{-1}$ and   $ \Omega_{\rm m} = 0.315$ \citep{Planck2020a} and the spectral index measurements given by the convection $S_\nu \propto \nu^{\alpha}$, where $S_\nu$ is the radio integrated flux density and $\alpha$ is the intrinsic source spectral index.

 
\section{Observations and Data Reduction}\label{s:datareduction}

\subsection{\emerlin{} SPARCS-North}
We obtained the SPARCS-North data from the \emerlin{} SPARCS Survey carried out on January 29 2018 at $1.5\,\mathrm{GHz}$ for a total of 24 hours (P.I. Wrigley). This observations employed a single-pointing strategy centred on R.A. $15^{h}33^{m}27^{s}$, Dec. $29^{\circ}12{\arcmin}40{\arcsec}$ covering an area of $\sim 15\arcmin \times 15\arcmin$ with a restoring beam size of $188 \times 170$\,mas. On-source integration time of $\sim 14\,$hr yielded an image with a central rms sensitivity of $\sim 10\,\microJybm$. The 52 phase centres used in our EVN+\emerlin{} survey were pre-selected from sources identified in this \emerlin{} survey above $\sim 6\sigma$, corresponding to sources with flux densities $>60\,\microJy$ \citep[Wrigley et al. in prep][]{}.

\subsection{EVN+ \emerlin{} observations}
\begin{figure*}
\centering
\includegraphics[width=1\linewidth, keepaspectratio=true]{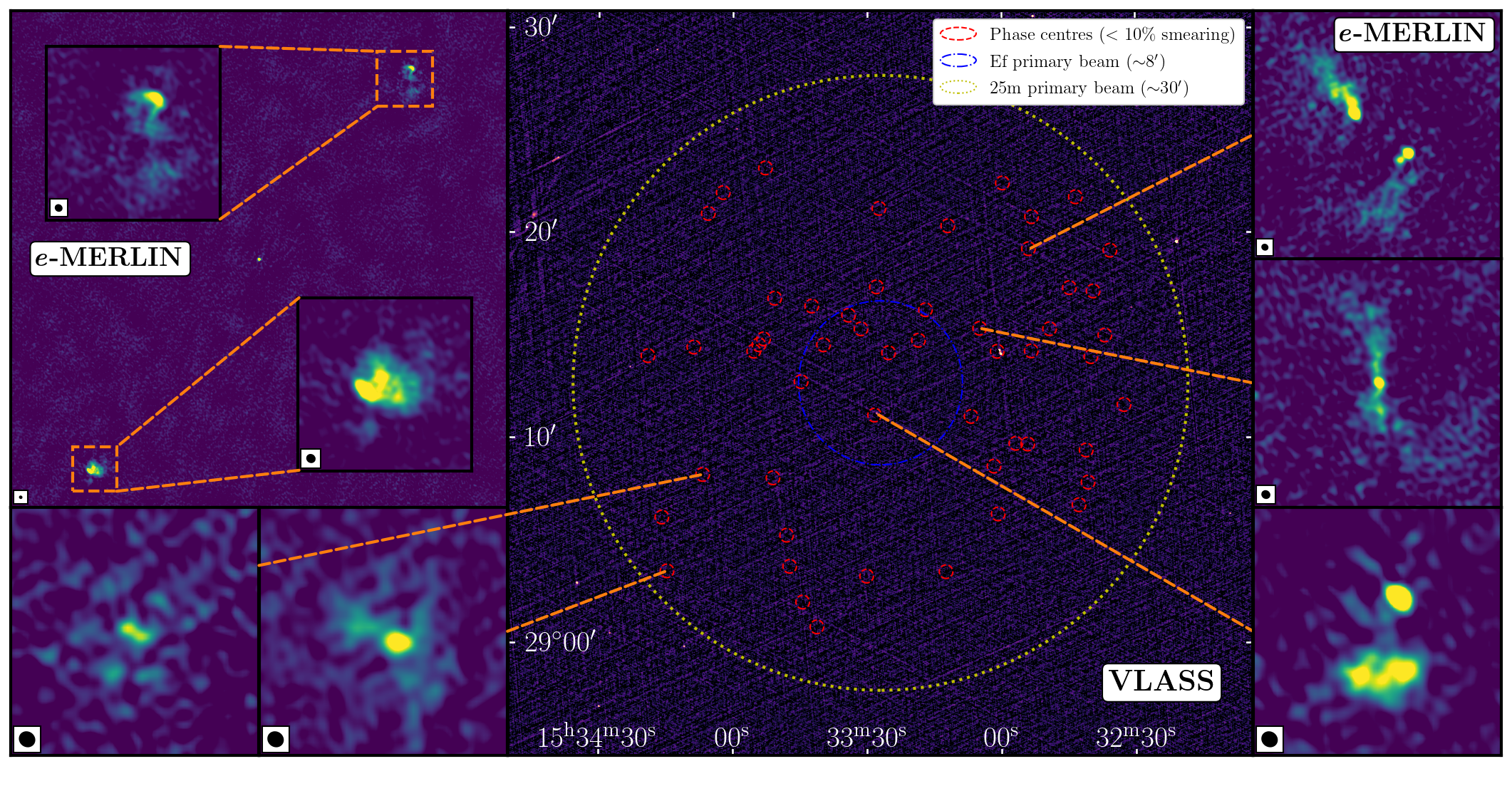}
\caption[]{The SPARCS field as seen by the VLASS and \emerlin{} surveys. The central plot comprises of the VLASS survey view of this field. Overlaid are the observed phase centres for this survey in the context of the Effelsberg (assuming an effective diameter of 78\,m) and the 25\,m primary beams of the EVN array. The extent of each phase centre corresponds to 10\% half-power beam width (HPBW) smearing, that is, assuming averaging to 2\,s integrations and $32\,\mathrm{kHz}$ channel width after the initial internal wide-field correlation and a maximum baseline of 10000\,km. Surrounding the VLASS image are postage stamp cut-outs of selected sources from the \emerlin{} pilot survey (r.m.s. $\sim 10\microJybm$), revealing the exquisite detail provided by the $0\farcs2$ high-resolution imaging.}
\label{fig:VLASS_eMERLIN_overlays}
\end{figure*}

The SPARCS-North continuum survey with the combined EVN + \emerlin{} array was carried out on November 3 2019 for a total of $19\,\mathrm{hrs}$ with a total on-source integration time of $\sim 14\,\mathrm{h}$. Participating EVN antennas are shown in Table 1 of \cite{Radcliffe2018}. The data was correlated at the Joint Institute for VLBI ERIC (JIVE) in Dwingeloo, the Netherlands. The observation setup frequency was 1594\,--\,1722\,MHz centred at $\sim$ 1658\,MHz, with a total bandwidth of 128\,MHz across 8 spectral windows and recorded at an aggregate bit rate of 1024\,Mbps with an integration time of $\mathrm{2\,s}$. We employed a single pointing observing strategy centred on R.A. $15^{h}33^{m}27^{s}$, Dec. $29^{\circ}12{\arcmin}40{\arcsec}$ using the multiple phase centre mode of the SFXC correlator \cite{Keimpema2015}. The multiple phase centre strategy was based on the 52 phase centres pre-selected from the \emerlin{} SPARCS survey with the furthest phase centre at $\sim 11\farcm6$ from the pointing centre. We aimed to observe all sources detected by \emerlin{} ($\geq60\mu$Jy, $6\sigma$), located all over the primary beam of the 25-m antennas (see Fig.~\ref{fig:VLASS_eMERLIN_overlays}) in the \emerlin{} array. Two sources, J1532+2919 and J1539+2744, were used for phase referencing. The source J1539+2744 (located $\sim$2$^{\circ}$ from the target field) was used as the primary phase (and delay) calibrator and was visited once every 30 minutes. This was further refined by visiting the secondary phase calibrator, J1532+2919 (located 0.27$^{\circ}$ from the target field) approximately every 7 minutes. The flux density for the primary calibrator is 198.0\,mJy and 42\,mJy for the secondary calibrator at C-band (from the Radio Fundamental Catalog, RFC, rfc2019d\footnote{\url{http://astrogeo.org/vlbi/solutions//rfc_2019d}}). The sources 4C39.25 (observed for $6\,$mins) and J1642+3948 (observed for $6\,$mins and revisited every four hours) were used as fringe finders.

\subsection{VLASS SPARCS-North}
We obtained the SPARCS-North images from the NRAO VLA Sky Survey (VLASS) archives. The VLASS is an all-sky Karl Jansky VLA 2-4 GHz survey covering the entire sky visible to the VLA upto a declination of $>-40^{\circ}$ at an angular resolution of $\sim 2\farcs5$ and an expected sensitivity rms of $70\,\mu$Jybeam$^{-1}$, across 3 epochs ($\sim 5500\,\mathrm{hr}$) over a cadence of 32 months from September 2017 - 2024 \citep{Lacy2020}. The VLASS covers a total of $33\,885 \mathrm{deg}^2$. We used the primary beam corrected image from the first epoch with a sensitivity rms of $\sim 120\,\microJybm$.

\subsection{Data processing}
The data was processed using the {\sc casa} based VLBI Pipeline \footnote{\url{https://github.com/jradcliffe5/VLBI_pipeline}}. Before calibration, the raw data was pre-processed by appending the system temperature ($T_\mathrm{sys}$) measured for each antenna to the FITS-IDI files. The effects of gravity on the shape of the antenna dishes as they move while tracking sources must also be accounted for. These distortions affect the gain of an antenna and are calculated based on the pointing direction of the antenna and the empirically observed properties. Additionally, the observatory flags  were translated into \textsc{casa} flags and finally the VLBI FITS-IDI (\textsc{AIPS} format) files were converted into a measurement set, (MS, \textsc{casa} format). The observatory flags were then applied to the MS and the correlated visibility amplitudes were scaled to a physically meaningful value using the system temperature measurements ($T_\mathrm{sys}$). We also carried out the total electronic content (TEC) correction for the ionosphere-induced dispersive delays. Automated flagging of RFI was conducted using the AOFlagger software \citep{Offringa2012} and any remaining bad data was flagged off manually.
 
The instrumental delays, which show up as jumps in phase between the spectral windows (spws) due to signal delays at each antenna, were removed by solving for the phase and delays on a 1 minute integration of the two fringe finders 4C39.25, J1641+3948 and the primary calibrator J1539+2744. The fringe-rates, that is, the derivative of the phase error over time, were set to zero to avoid extrapolating their effects across the entire experiment. The spws were then merged to increase the signal-to-noise (S/N) and further fringe fitting carried on the primary phase calibrator and the two fringe finders to remove group delays. The solution interval was set to $60\,\mathrm{s}$ such that it gave enough S/N to get good solutions, but short enough in order to track the changes in the delays, rates and phases. After the time-dependent delays were corrected for, the phase and rates were then calibrated and applied to the data. Bandpass calibration was then applied using the task {\sc bandpass} where the two fringe finders and the primary phase calibrator were used.

To correct for the phase errors due to the atmosphere, we used a bootstrapping approach using the primary phase calibrator, J1539+2744 (198\,mJy), located at  $2^{\circ}$ from the target field. This was then refined using the fainter (40\,mJy) secondary phase calibrator J1532+2919, located closer to the target field at $0.27^\circ$. The primary phase calibrator underwent two rounds of phase only calibration at a solution interval of 60\,s and 3 rounds of amplitude and phase calibration. These solutions were then applied to the secondary phase calibrator. The secondary phase calibrator then underwent 5 rounds of phase only calibration at a solution interval of 60\,s. The criterion for phase calibration selection is that the source should be located near the target and therefore the solutions derived over time for the phase calibrator are approximately correct for the target source, since the signal path through the atmosphere is similar for both sources. These solutions were then applied to a single phase centre to ensure that the corrections had been applied properly. The flagging tables and the calibration solutions derived for this single sub-field were then applied to all the remaining 51 phase centers.

\subsection{Primary beam corrections}{\label{PBC}}
Heterogenous arrays such as the EVN and \emerlin{} have different receptors and different dish diameters which result in each station having a different primary beam response. This manifests as a direction-dependent amplitude error due to differing attenuations of the antenna responses \citep[e.g.,][Radcliffe et al. in prep]{CottonMauch2021}. We derived corrections for the primary beam response of each antenna in the EVN+\emerlin{} array for each of the 52 phase centres based on the primary beam correction equation:

\begin{equation}
    P{(\theta)} = P{(0)}\mathrm{exp \left[-\frac{4ln(2)D_{\mathrm{eff}}^2}{\lambda^2}\theta^2\right]}, 
\end{equation}
where $P(\theta)$ is the corrected primary beam response, $P(0)$ is the normalised primary beam response = 1, ${D_\mathrm{eff}}$ is the effective dish diameter of the antenna, $\theta$ is the positional offset of the source from the pointing centre and $\lambda$ is the wavelength of the observation (see Table 1 of \citeauthor{Radcliffe2016}\citeyear{Radcliffe2018} for the ${D_\mathrm{eff}}$ values adopted).

\subsection{Imaging and source extraction}\label{s:sourceEXT}
\begin{figure*}
\centering
\includegraphics[width=\textwidth, keepaspectratio=True]{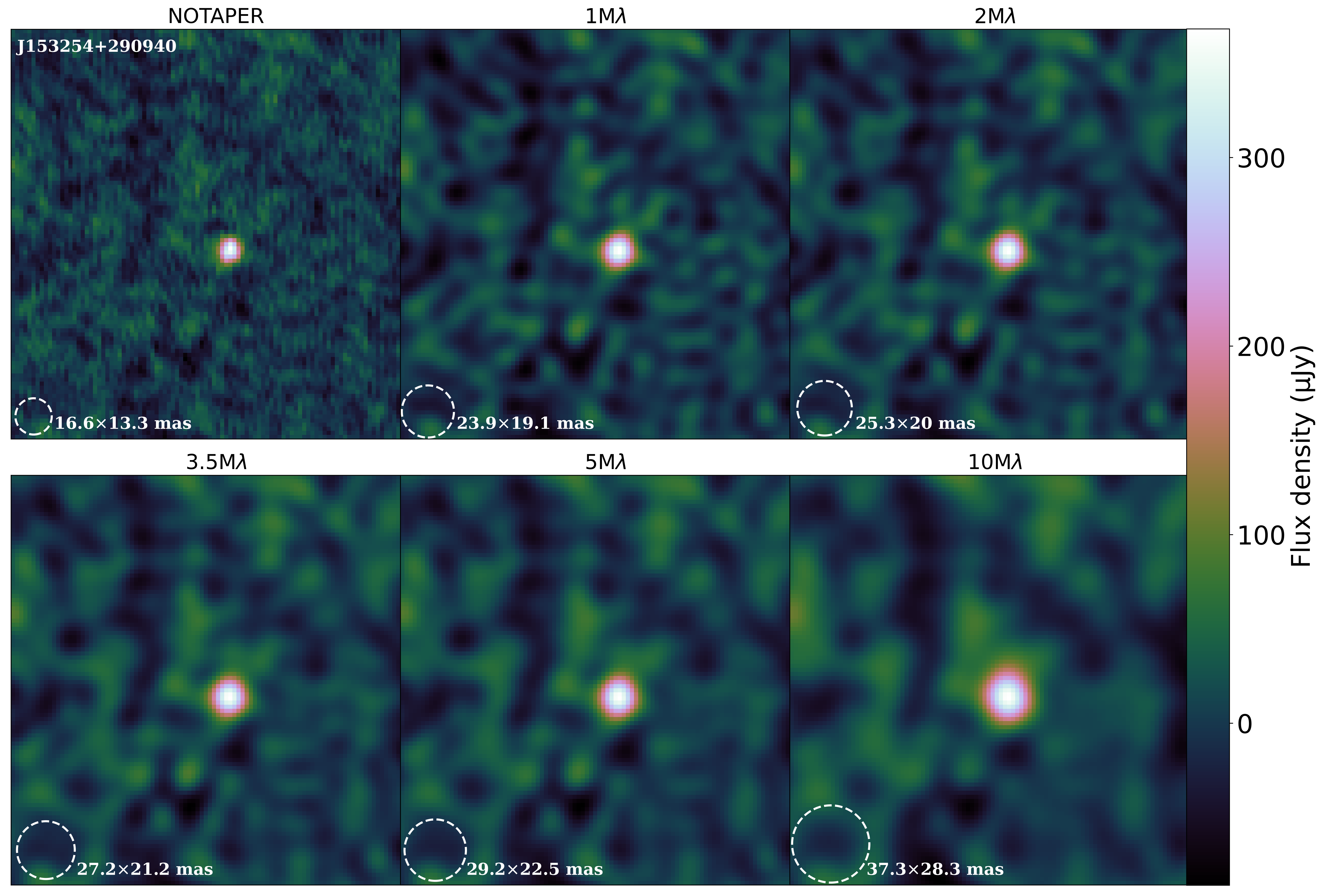}
\caption[]{We applied five \textit{uv}-tapers to the natural-weighted image at 1M$\lambda$, 2M$\lambda$, 3.5M$\lambda$, 5M$\lambda$ and 10M$\lambda$ which corresponded to an increase in the PSF from $\sim 13-40\,$mas for the source J153254+290940. The point source sensitivity decreased across the imaging at, 23.0$\microJybm$ for the natural-weighted only (NOTAPER) image, 27.9$\microJybm$ across the 1M$\lambda$ image, 28.4$\microJybm$ across the 2M$\lambda$ image, 29.2$\microJybm$ across the 3.5M$\lambda$ image, 30.0$\microJybm$ across the 5M$\lambda$ image and 32.4$\microJybm$ across the 10M$\lambda$ image. The 5M$\lambda$ image corresponded to the optimum detection point for point source sensitivity and surface brightness sensitivity and produced the highest peak brightness flux of $387\,\microJybm$ across the five \textit{uv}-tapers. Using this optimum taper value of $5M\lambda$, we produced a set of two images for each of the 52 phase centres, natural-weighted only (notaper image) and  natural-weighted with \textit{uv}-tapers applied at $5M\lambda$ (tapered image).}
\label{fig:tapers}
\end{figure*}

We used {\sc WSclean} to create $10\arcsec \times 10\arcsec$ images for each of the 52 phase centers. Natural-weighted images were created and 5 different outer uv-tapers applied. The weights at the highest \textit{uv}-distance were suppressed by a half-Gaussian kernel whose half-width half maximum (HWHM) was 1M$\lambda$, 2M$\lambda$, 3.5M$\lambda$, 5M$\lambda$ and 10M$\lambda$. with the corresponding resolution at FWHM ranging from $\sim 11 - 40\,\mathrm{mas}$ as shown in Fig.~\ref{fig:tapers}. Taking the weighting function along the uv-plane as $W(u,v)$ and sampling by a Gaussian: \begin{equation} W(u,v) = exp \left\{- \frac{(u^2+v^2)}{\mathrm{t}^2} \right\}\label{eq:tapering}\end{equation} where $\mathrm{t}^2$ is the tapering parameter (M$\lambda$ in this case). Tapering applies smoothing in the image plane by down-weighting the long baselines while degrading the angular resolution. This in turn decreases the point source sensitivity (PSS) while increasing the surface brightness sensitivity to extended radio emissions. We applied \textit{uv}-tapering to down-weight the contribution from the long baselines and to increase the surface brightness sensitivity on the shorter baselines, particularly along baselines with the Effelsberg and Lovell antennas. We note that the largest antennas requested only contribute to the inner regions of the image, enabling even deeper imaging at a central rms sensitivity  $\sim 6\microJybm$. The non-tapered image gave the highest PSS of $23.0\microJybm$, while tapering lowered the PSS. On the other hand, tapering increased the surface brightness sensitivity from 1200$\microJy$ in the non-tapered image to 1660$\microJy$ in the 10M$\lambda$ image. The restoring beam sizes, point source sensitivity and the surface brightness across the 5 tapers are shown in Table \ref{Tab:tapers}. The 5M$\lambda$ \textit{uv}-taper produced the optimum point source sensitivity and surface brightness sensitivity, beyond which, our observations became too sensitive to diffuse emissions and the point sources remained unresolved as demonstrated in Figure \ref{fig:tapers}. Therefore, we created two sets of images for each phase centre, a naturally-weighted image with no taper applied referred to as the notaper image and the naturally-weighted image with a 5M$\lambda$ taper applied referred to as the tapered image throughout this paper.

\begin{table}
    \caption{Radio properties of the source J153254+290940 at different tapers.}
    \label{Tab:tapers}
    \small
    \centering
    \begin{tabular}{lccc}
      
        \hline
        Taper & PSF & Point source & Surface brightness  \\
        & (mas) & Sensitivity ($\microJybm$) & Sensitivity ($\mathrm{mJy}$)  \\
        \hline
      Notaper & $16.6\times13.3$ & $23.0$ & $1.20$ \\ 
      
      1M$\lambda$ & $23.9\times19.1$ & $27.9$ & $1.40$ \\
      
      1M$\lambda$ & $25.3\times20.0$ & $28.4$  & $1.47$ \\
      
      2M$\lambda$ & $27.2\times21.2$ & $29.2$  & $1.52$ \\
      
      3.5M$\lambda$ & $29.2\times22.5$ & $30.0$ &  $1.60$ \\
      
      10M$\lambda$ & $37.3\times28.3$ & $32.4$ & $1.66$  \\

        \hline

\end{tabular}
\end{table}

For source extraction, we used the \textsc{PYBDSF} package \citep{Mohan2015} to determine the peak brightness and flux densities of our VLBI detections. \textsc{PYBDSF} computes basic statistics such as the image parameters, creates background rms and mean images, identifies islands of contiguous source emission whose peak brightness are above a given S/N threshold. The islands are then fitted with multiple Gaussian components in order to minimize the residuals with respect to the background rms, where the Gaussians within a given island are then grouped into discrete sources. We searched for everything above 5-sigma and filtered the catalogue to our detection thresholds. To eliminate any possibility of having random noise peak brightness exceeding this threshold, we applied the criteria used by \cite{Middelberg2013} to determine true source detections. Briefly, we estimated the number of independent resolution elements, $N$, based on Eq. (3) in \cite{Hales2012} combined with Eq. (6) in \cite{Middelberg2013} which estimates the number of beams exceeding $5\sigma$ if the image noise is Gaussian: 

\begin{equation}
    N_{+5} = N \times \left[1 - \mathrm{erf}\left(5\sqrt{2}\right)\right]/2,
\end{equation}

\begin{figure}
\centering
\includegraphics[width=1\linewidth, keepaspectratio=true]{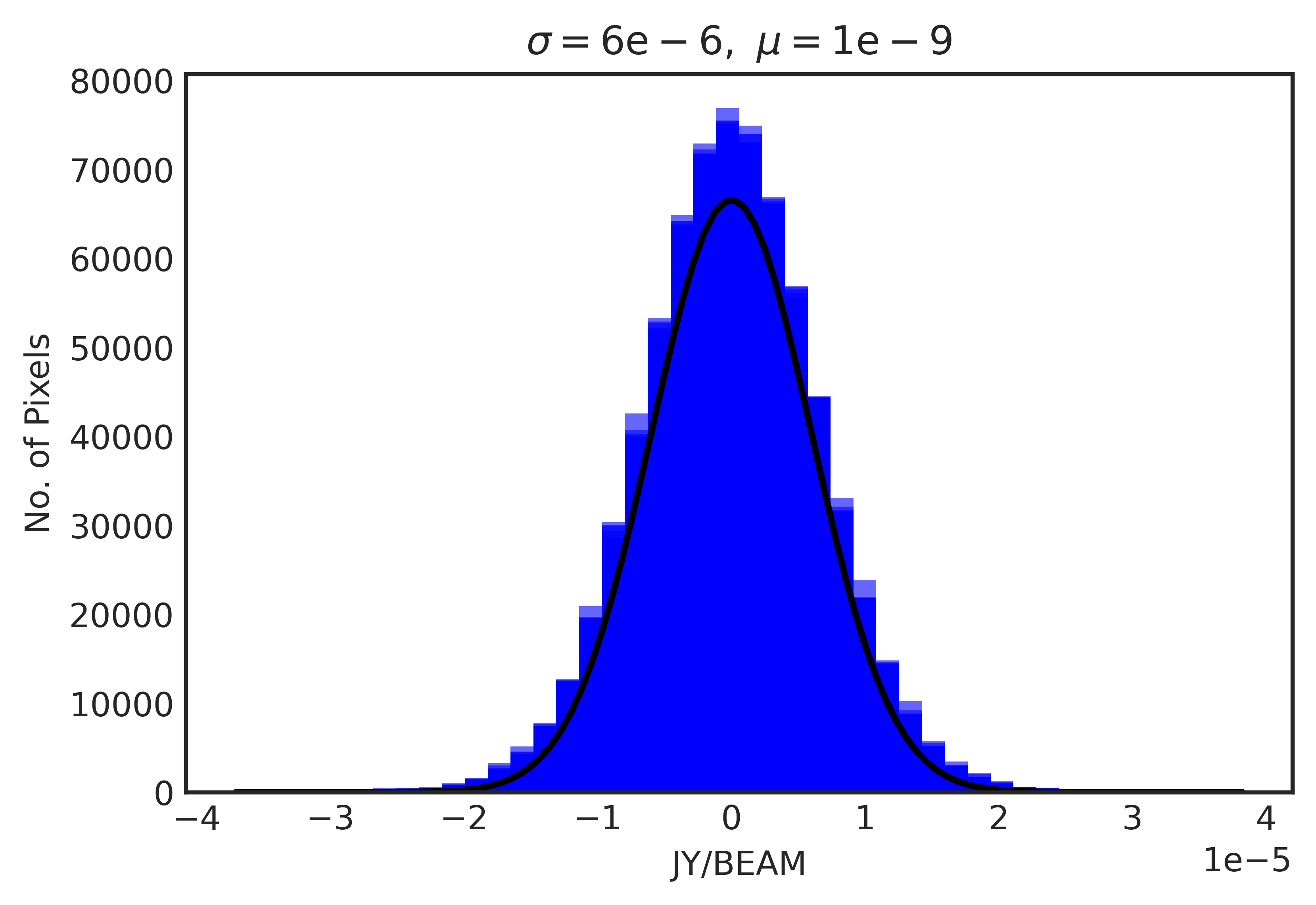}
\caption{Pixel distribution as a function of sensitivity showing a Gaussian distribution with $\sigma = 6\microJybm$ for the notaper image.}
\label{figure:pixeldist}
\end{figure}

We estimated about $\sim$ 600000 synthesized-beam areas in each 10\farcs0 by 10\farcs0 image centred on a single phase centre. The pixel distribution of each image is a Gaussian distribution, Fig.~\ref{figure:pixeldist}, and the probability of misidentifying random noise peak brightnesss as real detection was high. Assuming that the number of $>5\sigma$ peaks were of order of 10 within the \emerlin{} beam area and $5.5\sigma$ peaks were of order of 100 outside the \emerlin{} beam area , we applied the binomial probability distribution function \citep[see Eq. (7)][]{Middelberg2013} to calculate the true detection rates based on our two search-area criteria:

\begin{itemize}
    \item i) $>5.0\sigma$ for detections within the \emerlin{} beam area of radius $\sim 200\,\mathrm{mas}$, $\sim 0.25\%$ of the total area:
    \begin{equation}\label{eq:eMERLIN_Beam_Area}
    B(k=1 | p = 0.0025; n=10) = \binom{n}{k}p^k(1-p)^{n-k} = 0.02.
    \end{equation}
\end{itemize}

\begin{itemize}
    \item ii) $>5.5\sigma$ for detections outside the \emerlin{} beam area within $10\arcsec \times 10\arcsec$:
    \begin{equation}\label{eq:larger_image}
    B(k=1 | p = 0.01; n=100) = \binom{n}{k}p^k(1-p)^{n-k} = 0.0003.
    \end{equation}
\end{itemize}
The number of $>5\sigma$ noise peaks is smaller $<1$ across both criteria and further visual inspection ensured our $>5\sigma$ peaks were coincident with an \emerlin{} source. Any $>5.5\sigma$ peak detection not coincident with an \emerlin{} source was unlikely.

\begin{figure}
\centering
\includegraphics[width=1\linewidth, keepaspectratio=true]{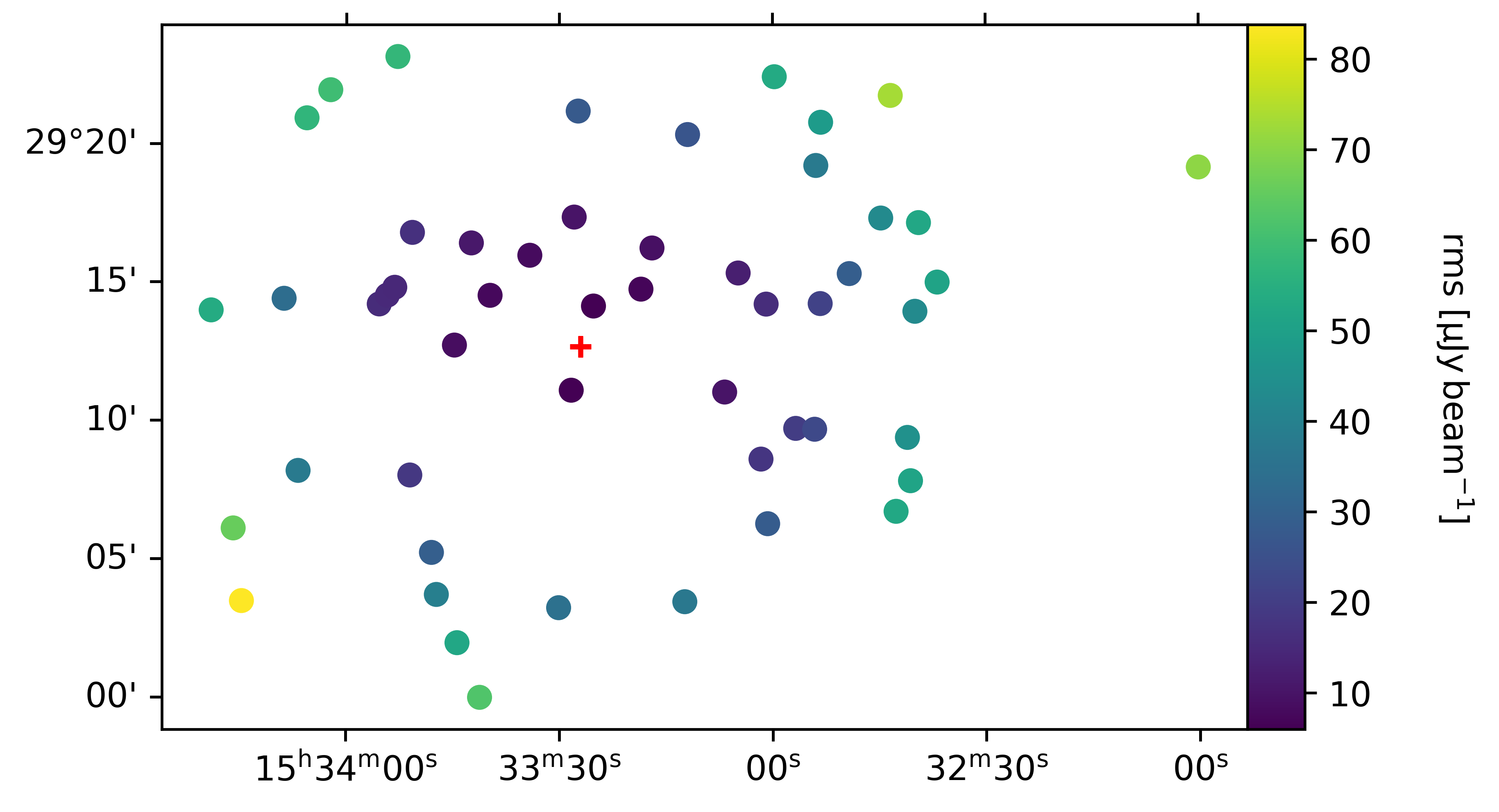}
\caption{The sensitivity map of the EVN+\emerlin{} SPARCS-North survey with a central rms sensitivity of $\sigma \sim 6\microJybm$. We computed the rms values of the 52 phase centres from the notaper images. The pointing centre is highlighted by the red cross.} 
\label{figure:rmsmap}
\end{figure}


\section{Results, Analysis and Discussion}\label{s:analysis}

\subsection{No taper imaging}
\begin{figure*}
\centering
\includegraphics[width=\textwidth, keepaspectratio=true]{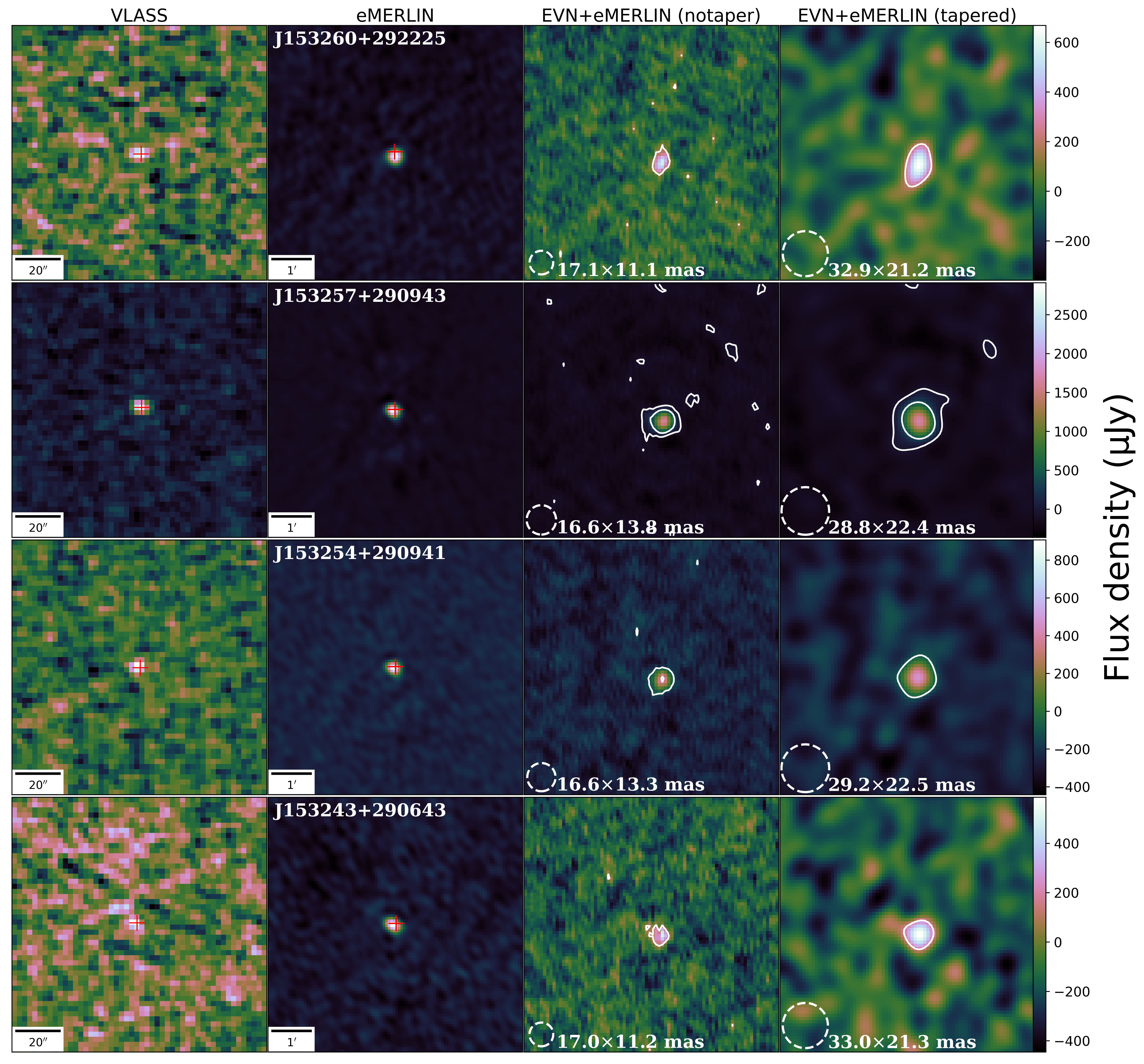}
\caption[]{Thumbnails of the SPARCS-North VLBI source sample showing coverage at different angular resolutions and sensitivity across the VLASS (column 1), \emerlin{} (column 2), EVN+\emerlin{} without tapers (notaper; column 3) and EVN+\emerlin{} with tapers (tapered; column 4). The notaper column shows the EVN+eMERLIN natural-weighted image with FWHM $\sim 17\times11\,$mas, while the tapered column gives the natural-weighted maps tapered at 5M$\lambda$ with FWHM $\sim 33\times21\,$mas. The notaper and tapered cutouts are 40\,pixel by 40\,pixel on a side. The notaper image offers the highest point source sensitivity (PSS), sufficient to reveal compact cores within the host galaxies as shown in column 3, while tapering increases the surface brightness sensitivity which reveals the appearance of sub-kpc jet-structures around the compact cores, possibly associated with either AGN jets or AGN embedded in star-forming regions as shown in column 4. The red cross indicates the VLBI position in VLASS and \emerlin{}. The flux density for the sources across \emerlin{}, EVN+\emerlin{} (notaper) and EVN+\emerlin{} (tapered) can be deduced from the colour bars. The restoring beam size is $2\farcs7 \times 2\farcs2$ for VLASS and $\sim0\farcs2 \times 0\farcs2$ for \emerlin{}. Three sources J153419+291359, J153406+292157 and J153342+291625 remain undetected in the VLASS survey.}
\label{fig:sample}
\end{figure*}

The pixel distribution as a function of sensitivity showed a Gaussian distribution, see Fig.~\ref{figure:pixeldist}, with the highest sensitivity recorded at $\sim6\microJybm$. This sensitivity decreased away from the pointing centre for the 52 phase centres as shown in Figure \ref{figure:rmsmap}. The notaper image provided the deepest imaging with a resolution at FWHM of $\sim 17 \times 11\,\mathrm{mas}$ and a central rms noise of $\sim6\microJybm$. Applying the binomial distribution function equations \ref{eq:eMERLIN_Beam_Area} and \ref{eq:larger_image} as described in Section \ref{s:sourceEXT}, 10 $>5\sigma$ VLBI peaks were detected within the \emerlin{} beam area of $0\farcs2$ by $0\farcs$2 and 20 $>5.5\sigma$ VLBI detections were recorded in the larger 10\farcs0 by 10\farcs0 area outside of the \emerlin{} beam area. We further carried out a visual inspection of these 30 VLBI detections using the \emerlin{} image. We confirmed all the 10/10 detections within the \emerlin{} beam area as true VLBI detections and 1/20 VLBI detection outside the \emerlin{} beam area with a high S/N $>350\sigma$ located at $5\farcs2$ from the phase centre. In total, we identified 11 VLBI candidates out of the 52 \emerlin{} sources. Figure \ref{fig:sample} shows the postage cutouts for these sources at different angular resolutions and sensitivity across the VLASS, \emerlin{}, EVN+\emerlin{}(notapers) and EVN+\emerlin{}(tapered).

\subsection{5M$\lambda$ tapered imaging}
The tapered image optimized both resolution and sensitivity, with a PSF of $\sim 33 \times 21\,\mathrm{mas}$ and a central r.m.s noise of $\sim7\microJybm$. Applying the binomial distribution function equations \ref{eq:eMERLIN_Beam_Area} and \ref{eq:larger_image} as described in section \ref{s:sourceEXT}, we detected a total of 21 potential VLBI detections,  with 7/21 VLBI detections at $>5\sigma$ constrained within the \emerlin{} area, and 14/21 $5.5\sigma$ VLBI detections recorded in the larger $10\farcs0$ by $10\farcs0$ image area outside the of \emerlin{} beam area. Applying visual inspection of our VLBI detections on the \emerlin{} image, we confirmed all the 7/7 $>5\sigma$ VLBI detections within the \emerlin{} beam area as true detections and 1/14 VLBI detection outside the \emerlin{} beam area with a high S/N $>300\sigma$ located at $5\farcs2$ from the phase centre. In total, we identified 8 VLBI candidates out of the 52 \emerlin{} sources. These 8 VLBI sources are also detected in the notaper image, see Table \ref{Tab:Sources}.

\begin{figure*}
\centering
\includegraphics[width=\textwidth, keepaspectratio=true]{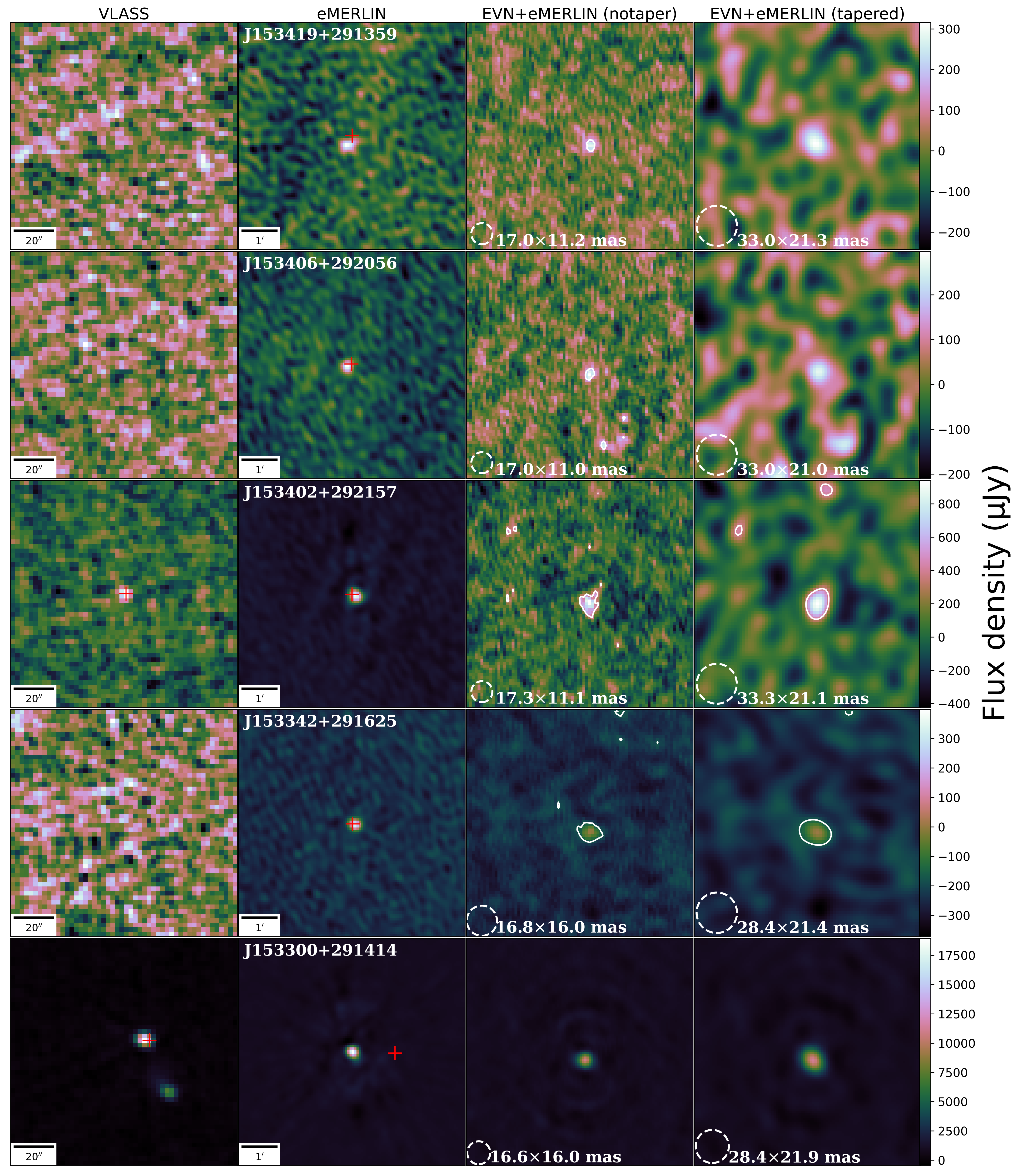}
\end{figure*}

\begin{landscape}
\begin{table}
    \caption{Radio properties of the VLBI sources.}
    \label{Tab:Sources}
    \small
    \centering
    \begin{tabular}{lccccccccccccc}
        \hline
        \multicolumn{4}{c}{ }&\multicolumn{5}{c}{notaper}&\multicolumn{5}{c}{tapered}\\
        \hline
        \textbf{Source ID}&\textbf{PC}&\textbf{RA(J2000)}&\textbf{DEC(J2000)}&\textbf{Beam}&\textbf{$I$}&\textbf{$P$}&\textbf{S/N}&\textbf{$R_\mathrm{notaper}$}&\textbf{Beam}&\textbf{$I$}&\textbf{$P$}&\textbf{S/N}&\textbf{$R_\mathrm{tapered}$}  \\
        & & & & (mas) & (\flux{}) & ($\microJy$/beam{}) & & & (mas) & (\flux{}) & ($\microJy$/beam{}) & &  \\
        \hline
        
         J153419+291359 & 402 & 15:34:18.9806 & +29:13:59.2224 & 17\x15 & 273\plus13 & 264\plus7.4 & 5.5 & 0.46 & x & x & x & x & x  \\
         
        J153406+292056 & 407 & 15:34:05.5352 & +29:20:55.5868  & 17\x11 & 197\plus52 & 307\plus29 & 5.2 & 0.93 & x & x & x & x & x  \\
        
        J153402+292157 & 408 & 15:34:02.2050 & +29:21:56.5011 & 17\x11 &  573\plus135 & 429\plus55 &  6.3 & 0.57 & 33\x21 & 465\plus102 & 629\plus69 & 7.3 & 0.40 \\
         
        J153354+291431 & 410 & 15:33:54.1994 & +29:14:31.4017 & 17\x15 & 268\plus17& 282\plus10 &18.4 & 0.45 & 28\x22 & 254.8\plus6.6 & 278\plus4 & 14.7 & 0.42   \\
         
        J153354+291431 & 410 & 15:33:54.1997 & +29:14:31.5338 & 17\x11 & 74\plus23 & 75.6\plus16 & 5.5 & 0.46 & x & x & x & x & x \\
         
        J153342+291625 & 419 & 15:33:42.3797 & +29:16:24.7116 & 17\x16 & 115\plus11 & 127.1\plus6.9 & 11.1 & 0.33 & 28\x21 & 126\plus18 & 117\plus9.9 & 9.2 & 0.36 \\
         
        J153300+291414 & 435 & 15:33:00 & +29:14:14 & -- & -- & -- & $>350$ & -- & -- &-- & -- & $>300$ & \\
                
        J153260+292225 & 437 & 15:32:59.7091 & +29:22:24.8132 & 17\x11& 511\plus43 & 492\plus24 & 8.8 & 0.77 & 33\x21 & 564\plus28 & 570\plus15 & 7.8 & 0.85 \\
         
        J153257+290943 & 438 & 15:32:56.7349 & +29:09:43.4550 & 17\x14 & 1328\plus67 & 1260\plus38 & 64.2 & 0.55 & 29\x22 & 1242\plus35 & 1280\plus21 & 49.4  & 0.51 \\
         
        J153254+290941 & 439 & 15:32:54.0822 & +29:09:40.7617 & 17\x13 & 378\plus24 & 369\plus13 & 15.7 & 0.55 & 29\x22 & 367\plus10 & 394.9\plus6.3 & 13.0 & 0.53   \\
         
        J153243+290643 & 446 & 15:32:42.6489 & +29:06:42.6796 & 17\x11 & 409\plus10 & 360\plus53 & 5.9 & 0.88 & 33\x21 & 407\plus38 & 504\plus25 & 6.7 & 0.87  \\

        \hline
        
\vspace{-0.2cm}\\    
\multicolumn{14}{l}{\footnotesize \textbf {Notes:} PC is the  source phase centre number. The cross x highlight non-detection in the tapered image. The source PC = 435 showed significant image artefacts around a bright } \\ 
\multicolumn{14}{l}{source and thus difficult to deduce the flux density of the source.}\\        
    \end{tabular}

\end{table}
\end{landscape}

\subsection{Astrometry}
\begin{figure}
\centering
\includegraphics[width=1\linewidth, keepaspectratio=true]{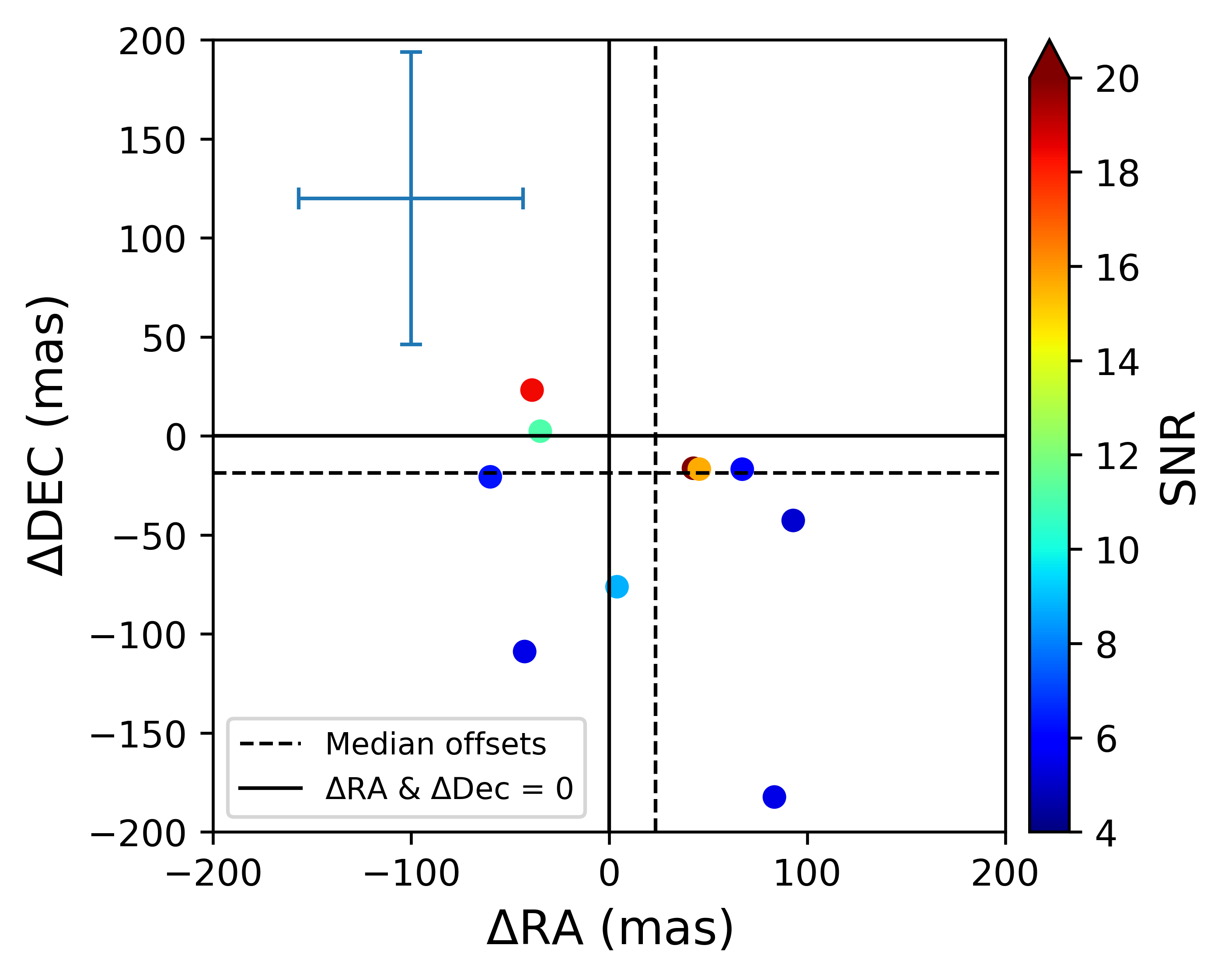}
\caption{The VLBI astrometric offsets, $\Delta$RA and $\Delta$Dec, between the EVN+eMERLIN and the \emerlin{} SPARCS-north surveys. The solid black lines indicate $\Delta$RA = 0 and $\Delta$Dec = 0 mas while the broken lines indicate the median astrometric offsets $23.4\,$mas in $\Delta$RA and $-18.7\,$mas in $\Delta$Dec. The cross represents the typical error on each source. The VLBI S/N of the sources can be deduced from the colour bar.}
\label{fig:SPARCSoffsets}
\end{figure}

10/11 sources were detected within a radius of $\sim 0\farcs2$ of an \emerlin{} source in both the notaper and tapered images. The remaining source, J153300+291414, was detected $5\farcs2$ from the \emerlin{} counterpart. The VLBI detections show a median astrometric offset of $23.4\,$mas in $\Delta$RA and $-18.7\,$mas in $\Delta$Dec between this EVN+eMERLIN survey and the \emerlin{} survey, shown in Fig. \ref{fig:SPARCSoffsets}.

\subsection{Tapered image non-detections}\label{sec:interestinfsources}
3/11 VLBI detections in the notaper image remained undetected in the tapered image, with 2/3 sources also remaining undetected in VLASS.

We define the flux density ratios, $R$ between the VLBI and \emerlin{} as:
$R_\mathrm{notaper} = S_\mathrm{i,notaper} / S_\mathrm{i, e-MERLIN}$ for the notaper, and $R_\mathrm{tapered} = S_\mathrm{i,taper} / S_\mathrm{i, e-MERLIN}$ for the tapered image. The $R$ values for the 11 sources are presented in Table \ref{Tab:Sources}.

\subsubsection{J153419+291359}
The source is detected in the notaper image at a 5.5$\sigma$ but remains undetected in the tapered image due to the increased noise levels. The rms in the tapered image is higher at 73\,$\microJybm$ compared to the notaper rms at 53\,$\microJybm$. The source has an integrated flux density of $273\pm13\,\mu\mathrm{Jy}$ and a peak brightness of $264\pm7.4\,\mu\mathrm{Jybeam^{-1}}$ in VLBI and $594.8\pm5.7\,\microJy$ and $550.1\pm3.1\,\microJybm$ in \emerlin{} with $R_\mathrm{notaper} = 0.46$. The source remains undetected in VLASS, with the \emerlin{} peak brightness $\sim 5\times$ the VLASS r.m.s. If we assume a central frequency of 1600\,MHz for the VLBI and the \emerlin{} observations and 3\,GHz for VLASS we can calculate the ratio depending on the spectral index of the source. Typical extra-galactic sources have spectral indices of about -0.77 \citep[e.g.,][]{Kellermann1963}, which means that for any source, the VLASS peak should be about 50\% of the VLBI peak.


\subsubsection{J153406+292056}
The source is detected in the notaper image only at $5.2\sigma$. The VLBI integrated flux and peak brightness are $197\pm52\,\microJy$ and $307\pm29\,\microJybm$ while the \emerlin{} integrated flux and peak brightness are $211\pm23\microJy$ and 301\plus16$\,\microJybm$ with $R\mathrm{_{notaper}} = 0.9$. The non-detection in the tapered image can be attributed to the fact that this is a very faint source and the noise levels in the tapered image ($\sim 77.5\,\microJybm$) are higher than in the notaper image ($\sim 57\,\microJybm$).  The source is also remains undetected in the VLASS image, with the \emerlin{} peak brightness 2 times the VLASS noise level. This non-detection in VLASS could be attributed to AGN variability, see Section \ref{sec:Varibiality}.

\subsubsection{J153354+291431}
The source shows binary VLBI compact cores in the notaper image and a single core in the tapered image (see Fig.\ref{fig:source410}). In the notaper image, the brighter core is detected at S/N $\sim 18.4$, while the second fainter core is detected at $\sim 5.5\sigma$. In the tapered image, the brighter core is detected at $\sim 14.7\sigma$, while the second fainter core remains undetected. The brightest VLBI core has a flux density of 268.0\plus 17.0\flux{} and peak brightness of 282.0\plus10.0$\,\microJybm${} in the notaper image with $R\mathrm{_{notaper}} = 0.45$, and a flux density of 254.8\plus6.6\flux{} and peak brightness of 278.0\plus4.0$\,\microJybm${} in the tapered image with $R\mathrm{_{tapered}}= 0.42$. The fainter VLBI core has $R\mathrm{_{notaper}} = 0.46 $ with a flux density of 74\plus{13}\flux{} and 75.6\plus{7.4}$\microJybm${} in the notaper image. See section \ref{sec:410} for further analysis and discussion. 

\subsubsection{J153342+291625}
The source has a detection in both the notaper and the tapered images  but remains undetected in VLASS. It was detected at S/N $11.1\sigma$ in the notaper image and $9.2\sigma$ in the tapered image. The notaper VLBI integrated flux density is $117\pm{11}\,\mu\mathrm{Jy}$ and a peak brightness of $127.1\pm6.9\,\microJybm$ with $R\mathrm{_{notaper}} = 0.33$. The integrated flux and the peak brightness flux across the tapered image are 126\plus{18}$\,\mu\mathrm{Jy}$ and $122.4\,\microJybm$ respectively, with $R\mathrm{_{tapered}} = 0.33$. The \emerlin{} has a flux density of 352.5\plus6.4\,$\microJy$ and a peak brightness flux of 320.4\plus2.5\,$\microJybm$ with the \emerlin{} peak brightness $\sim ~3\times$ the VLASS r.m.s.  

\subsection{Multi-resolution flux density comparison}
Excluding the two sources, J153257+290943 and J153300+291414, our VLBI sources have integrated flux and peak brightness ranging from 75-665\flux{} and 75-492$\microJybm$ in the notaper imaging, and 116-465\flux{} and 122-629$\microJybm${} across the tapered imaging. The median $R\mathrm{_{notaper}} = 0.51$ and $R\mathrm{_{tapered}} = 0.55$ implying that we recovered $\sim 55\%$ of the \emerlin{} flux density across the tapered imaging and $\sim 51\%$ across the notaper imaging. The source J153257+290943 has a flux density and peak brightness of 1328\flux{} and 1260$\microJybm$ in the notaper image and 1242\flux{} and 1280$\microJybm${} in the tapered image, while the source J153300+291414 shows some strong artefacts and thus the flux density could not be deduced. Only 3/11 sources are resolved in the notaper image, whereas all sources remained unresolved in the tapered image. This implies that resolutions at FWHM $\sim 30\,\mathrm{mas}$, the survey becomes more sensitive to  diffuse emissions than compact emissions.

The three VLBI sources, J153419+291359, J153406+292056 and J153354+291431, non-detected in the tapered image have a S/N $\sim 5\sigma$ across the notaper imaging. The tapered rms is higher than the notaper rms for all 3 sources. This implies that the higher PSS achieved in the notaper imaging increases sensitivity towards very faint compact regions, which are otherwise missed out at intermediate spatial scales. Assuming $z\sim 1.25$, the $11-33\,\mathrm{mas}$ recovered in this survey correspond to spatial scales of $\sim 9 - 280\,\mathrm{pc}$. Furthermore, most of our VLBI sources are point-like and show no jet structure in the notaper image, while all sources show emergence of either one-sided or two-sided jet structures at $\sim$sub-kpc sizes across the tapered image, see Figure~\ref{fig:sample}. This could indicate that at spatial scales of  $\sim 0.28\,$kpc, our survey becomes more sensitive to diffuse emission, associated with either AGN with jets, star formation or both, while anything below $<0.28\,$kpc can be classified as pure AGN. This cut-off spatial scales could potentially pinpoint the transition point from sources dominated by compact emissions purely associated with AGN, to regions dominated by extended diffuse emissions, and thus isolating AGN contribution from star formation calculations. 



\subsection{Compact sources vs. extended radio sources}\label{sec:emerlinsources}
\begin{figure}
\centering
\includegraphics[width=1\linewidth, keepaspectratio=true]{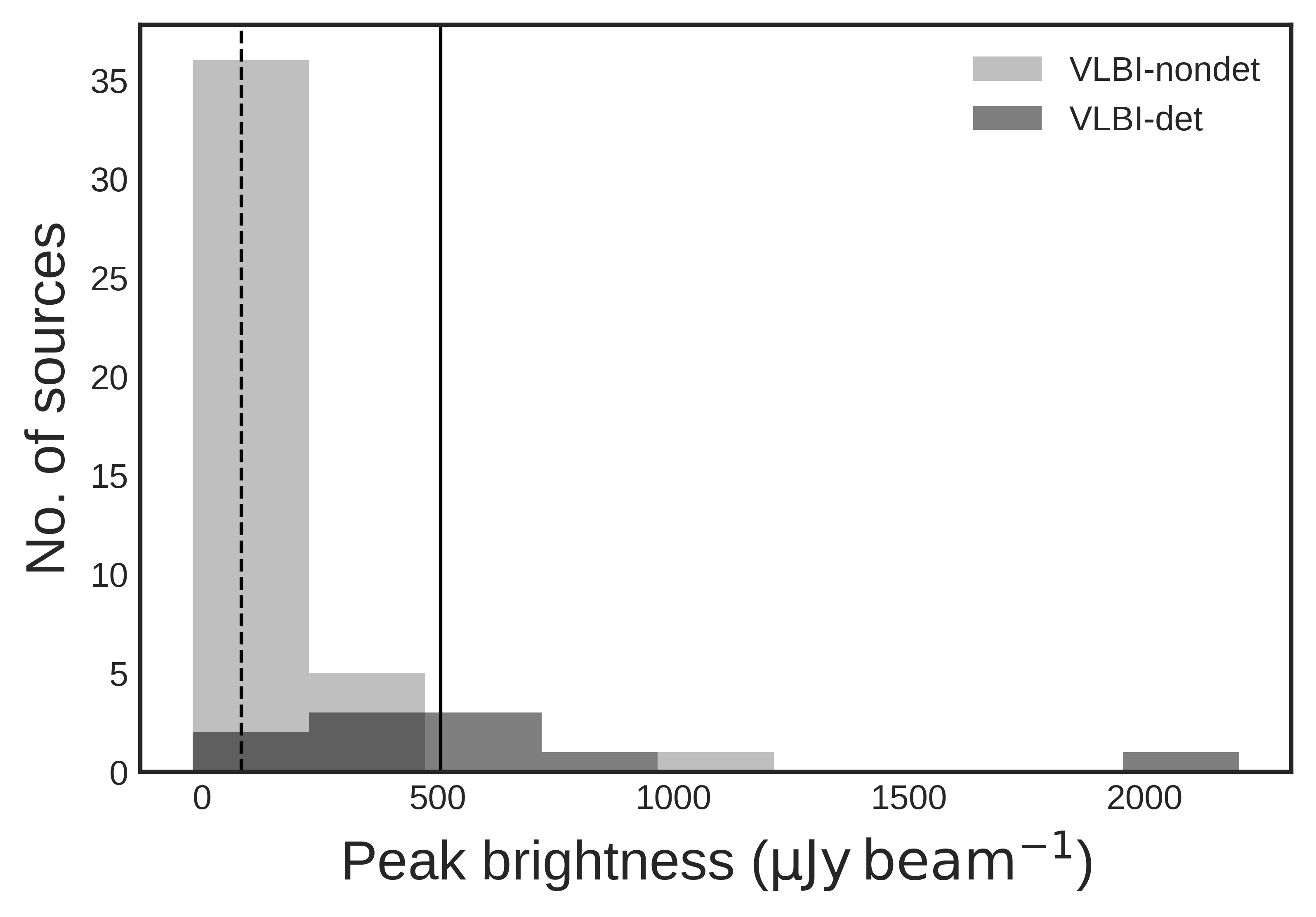}
\caption{Peak brightness distribution for the 52 \emerlin{} sources used as the phase centres in this survey.The region shaded in grey highlights the peak flux distribution for the 41/52 \emerlin{} sources that remain undetected (VLBI-nondet) in this survey, while the region highlighted in black shows the peak flux distribution for the 11/52 \emerlin{} sources detected (VLBI-det). The vertical lines give the median peak brightness flux, the dotted line is 83.25$\mathrm{\mu Jy\,beam^{-1}}$ for the non-detections and 506.25$\mathrm{\mu Jy\,beam^{-1}}$ for the VLBI detections.}
\label{fig:peakdistribution}
\end{figure}

In this VLBI survey, we detected all the 11 compact \emerlin{} sources, while the remaining 41/52 non-detected sources showed diffuse and extended radio structures in \emerlin{}. This represents a 100\% VLBI detection rate for the compact \emerlin{} sources. The 41/52 VLBI non-detections have integrated flux densities across the \emerlin{} image of $\sim$ 60--2550\,$\microJy$ with a median of $143.75\,\microJy$, while the 11/52 VLBI detections have 159.10--7500$\,\microJy$ with a median of $597.10\,\microJy$, respectively. The VLBI non-detections have peak brightness of $6.70 - 1039\,\microJybm$ with a median of $83.25\,\microJybm$, while the 11/52 VLBI detections have peak brightness of 113.40 -- $2171\,\microJybm$ with a median of $506.25\,\microJybm$. Figure~\ref{fig:peakdistribution} shows the peak brightness ($\microJybm$) distribution for the VLBI detected and non-detected sources in the \emerlin. We expect an \emerlin{} source to have a peak brightness $\sim 5\times$ the local rms ($>300\,\microJybm$) for a detection in VLBI, $\sim 82\%$ of the VLBI detections show peak brightness distribution $> 300\,\microJybm$. Comparing both the flux density and peak brightness for the VLBI non-detections and VLBI detections and as expected, the VLBI is biased towards very compact cores of the fainter sources with higher peak brightness while the low surface brightness sources remain undetected. 

Overall, this represents a VLBI detection rate of $\sim 21\%$ at $1\sigma$ sensitivity of $6\,\microJybm$. This VLBI detection rate is in agreement with other VLBI detection fractions such as the $20\pm1\%$ at $1\sigma$ sensitivity of $10\,\microJybm$ in the VLBA COSMOS survey \citep{Herrera2017}, $20\pm0.3\%$ at $1\sigma$ sensitivity of $60\,\microJybm$ in the mJIVE project \citep{Deller2014} and the $20^{+5}_{-4}\%$ at $1\sigma$ sensitivity of $55\,\microJybm$ in the Chandra deep field south (CDFS) survey.


\subsection{AGN variability}\label{sec:Varibiality}
3/11 VLBI sources, J153419+291359, J153406+292056 and J153342+291625 remain undetected in VLASS, see section \ref{sec:interestinfsources}. For a VLBI source to be detected in VLASS, we assume that the peak brightness must be $>5\times$ the local rms in VLASS. With a rms sensitivity of $\approx 120\,\microJybm$, the source peak brightness must be $>600\,\microJybm$ for a detection in VLASS. All the 8/11 VLBI sources with VLASS counterparts have peak brightness $>600\,\microJybm$ in VLASS. Since all 11 VLBI sources are detected in the \emerlin{} image, this non-detection in VLASS can be attributed to either variable compact cores or steep spectral indices. Attributing this non-detection to variable compact cores, this could be due to the increase in fluxes as both the VLASS and \emerlin{} observations were carried out at different times, with the \emerlin{} observations taking place $\sim 4$ months after the VLASS observations. For example, when comparing the first epoch of VLASS (2017 - 2019) to the FIRST survey (1993-2011), \cite{Nyland2020} show almost a ten-fold flux increase in flux for sources co-located with a known AGN and their results indicate that the slow transient radio sky is dominated by emission mostly attributable to AGN variability, with variability timescales of a $\sim$ few months. Furthermore, variability studies of the faint $\umu$Jy radio source population using high radio resolutions and multiwavelength observations have revealed that almost all sources showing variability are AGN \citep[e.g.,][]{Radcliffe2019,Sarbadhicary2021}. $\sim$ 3 per cent of our sources exhibiting potential variability, is in agreement with the short-term variability study of the variable $\umu$Jy radio sky of \cite{Radcliffe2019} with 3 per cent of their sources exhibiting AGN-related variability. Furthermore, in the CHILES VERDES study of radio variability in the COSMOS field, \cite{Sarbadhicary2021} showed that no star-forming galaxies exhibited any significant variability and therefore attributed all their variability to AGN. From our survey, we further note that a majority (2/3) of these sources potentially exhibiting variability are faint VLBI sources detected at $5.5\sigma$ and $5.2\sigma$. At the sensitivity and spatial scales recovered in this survey, we expect to capture pure AGN emissions and therefore, useful in further probing variability as a powerful diagnostic tool for AGN classification. We aim to carry out a similar survey with the \emerlin{} and the EVN+\emerlin{} probing both short and long-term variability in these radio source populations.

\section{Serendipitous discovery?}\label{sec:410}
\begin{figure*}
\centering
\includegraphics[width=0.78\textwidth, keepaspectratio=true]{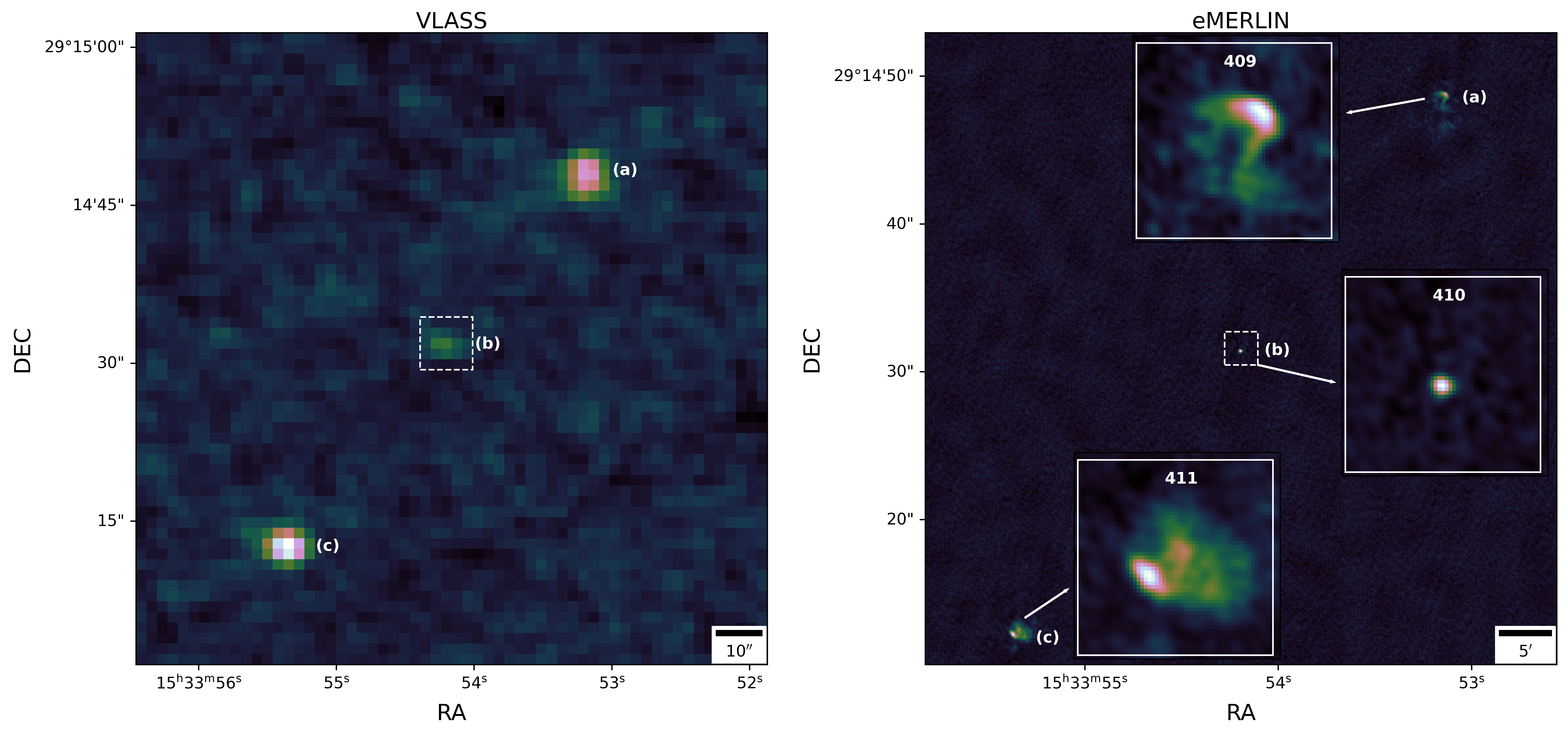}
\includegraphics[width=0.18\textwidth, keepaspectratio=true]{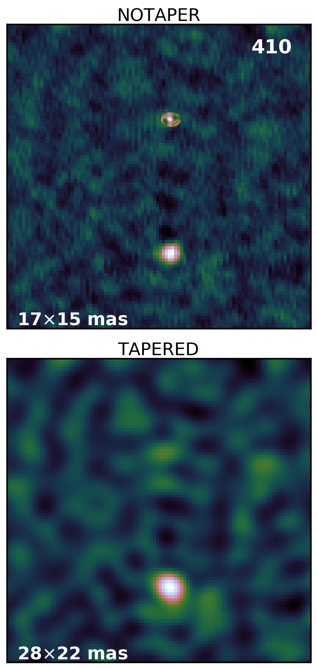}
\caption[]{The source J153354+291431 shows two bright compact cores on the VLBI image (NOTAPER and TAPERED), which are located within a double-lobed structure indicating the presence of highly collimated radio jet structures at $\sim$ galactic scales in both the VLASS and the \emerlin{} images. The VLASS image shows 3 unresolved sources at (a), (b)  and (c) which correspond to 3 \emerlin{} sources (PC 409, 410 and 411, respectively). Whereas the source at point (b) remains faint and compact in \emerlin{}, the sources at points (a) and (c) are resolved out revealing extended radio emissions with edge-brightened outer lobes showing some hot spots. The faint component (b) is coincident with the centre of the host galaxy in both VLASS and \emerlin{}. On the other hand, VLBI completely resolves out these outer lobes and hotspots, but only partially resolves the faint component (b) to reveal two bright compact cores in the notaper image. These two VLBI cores are separated by $\sim 130\,$mas corresponding to $\sim 1.1\,$kpc, which is well within the 2.5\,kpc orbital separations between paired supermassive black holes proposed by \cite{Burke2011}. The fainter (northern core) is detected at $\sim 5.5\sigma$ and the brightest (southern core) is detected at $\mathrm{\sim 18.4\sigma}$. In the tapered image, the fainter northern core remains undetected, whereas the brighter southern core is detected at $14.7\sigma$. The distance between the two lobes in \emerlin{} is $\sim 46$\farcs, translating to a linear size of  of $\sim 0.4\,$Mpc, assuming $z \sim 1.25$.}
\label{fig:source410}
\end{figure*}

The source J153354+291431 shows two compact VLBI cores in the notaper image and one compact VLBI core in the tapered image that are located within a double-lobed structure in both the VLASS and the \emerlin{} images as shown in Figure \ref{fig:source410}. The fainter (northern core) is detected at $\sim 5.5\sigma$ in the notaper image only, while the brightest (southern core) is detected at $\mathrm{\sim 18.4\,\,and\,\,14.7\sigma}$ across both the notaper and tapered images, respectively. The double-lobed structures are well aligned in the north-east south-west direction across both images, indicating the presence of highly collimated radio jet structures at $\sim$ galactic scales. In the \emerlin{} imaging, the double-lobed structure reveal extended radio emissions with edge-brightened lobes showing some hot spots. The two lobes have a separation of $\sim 46\arcsec$ in \emerlin{}, translating to a linear size of $\sim 0.4\,$Mpc, assuming $z \sim 1.25$. The unresolved fainter middle component, component (b) in Figure \ref{fig:source410}, is coincident with the centre of the host galaxy in both VLASS and \emerlin{}. VLBI completely resolves out the outer lobes and partly resolves the fainter middle component to reveal two bright compact VLBI cores separated by a distance of $\sim 130\,$mas ($\sim 1.1\,$kpc, assuming $z \sim 1.25$) in the notaper image. This separation distance is well within the 2.5\,kpc orbital separations between paired supermassive black holes proposed by \cite{Burke2011} and $1\,$pc -- $100\,$kpc of \cite{DeRosa2019}. 

The VLASS survey is designed to identify such dual AGN at separations $<7\,$kpc and binary supermassive black holes within ongoing mergers \citep{Burke2014,Burke2018}. The VLASS survey detected all the three unresolved components within this system, components (a), (b) and (c) in Figure \ref{fig:source410}, which was useful in our host galaxy identification. The \emerlin{} observations on the other hand resolve out the extended radio emissions revealing edge-brightened outer lobes with hotspots and are therefore important for probing active radio galaxies in which accretion onto the central SMBH generates relativistic jets resulting in large-scale environments as is the case with this source \citep[e.g.,][]{Mingo2019,Hardcastle2019,Croston2019}. VLBI observations completely resolve out all the extended emissions associated with the host galaxy, and partly resolves the faint middle component revealing the two compact cores at the centre of this galaxy. This shows that, even at the smallest orbital separations, mas resolutions provided by VLBI are perfect for exploring and identifying dual AGN or binary SMBHs, particularly within merging galaxy systems which are otherwise difficult to identify as double systems as evidenced here. This amplifies the importance of VLBI in the study of binary systems in post-merger galaxy systems \citep[e.g.,][]{Wassenhove2012}, stalling in binary black holes in the intermediate stages, besides providing statistical estimates of the rate of SMBH inspiral at various phases \citep[e.g.,][]{Burke2011,Arca-Sedda2019,Mingarelli2019}. 

This source is a possible paired SMBH candidate, in a massive ($\sim 0.4\,$Mpc in size) radio loud AGN (RLAGN) system. This possible serendipitous discovery is unique but not entirely unexpected, as shown in similar high-resolution VLBI discoveries. For example, \citet{Spingola2019}, found two unexpected gravitational lensing candidates within the wide-field mJIVE-20 survey \citep{Deller2014}. These two candidates can be used to constrain the sub-halo mass function which in turn influences the constrains of the the cosmological parameters and galaxy formation models \citep{Spingola2018}. \citet{Herrera2017} also finds two unexpected binary SMBH candidates within the VLBA-COSMOS survey. We aim to propose an even deeper follow-up survey to confirm the nature of this source \citep[following e.g.,][]{Deane2014,Breiding2022} and to further probe phenomena such as merger-induced SBMH growth with simulations predicting a peak in the growth rate at $1 \-- 10\,$kpc \citep{Wassenhove2012} and $0.1\--2\,$kpc \citep{Blecha2013}.

\section{Conclusions}\label{s:conclusions}
We have presented a direct view of AGN in the host galaxies and the physical interplay between SF and the nuclear activity through a multi-resolution $\sim 10-100\,\mathrm{mas}$ high dynamic (9\,pc--0.3\,kpc at $z\sim 1.25$) view of a moderate sample of galaxies. We detect a total of 11 VLBI sources at $>5.0\sigma$ from 52 potential sources based on the \emerlin{} survey of the SPARCS-North field. This EVN+\emerlin{} data provides the radio flux densities and morphology information required to isolate pure AGN from star-formation processes within individual galaxies. Despite the small sample size, this information reduces the systematic uncertainties by quantifying the AGN contributions to be subtracted from calculations of SF based on radio continuum. In principle, this survey permits the inclusion of galaxies with AGN hosts which would normally be excluded from such calculations, as evidenced by sources J153419+291359, J153406+292056 and J153354+291431. This is absolutely necessary if we are to obtain the tightest ever constraints on the SF measurements and to resolve out the total contribution of AGN to the cosmic SF history of the host galaxies \citep[e.g.,][]{Macfarlane2021}.

At spatial scales of $\sim9\,$pc, the sources show little to no jet structure whilst at $\sim 0.3\,$kpc one-sided and two-sided radio jets start to emerge on the same sources, indicating a possible transition from pure AGN emissions to AGN plus star-formation systems. The arcsecond spatial resolutions offered by the \emerlin{} and VLA are important in identifying the host galaxies within this faint ($\microJy$) radio source populations, the VLBI on the other hand has a selection bias towards high temperature brightness $>10^5\,$K and therefore sensitive to compact emissions mostly indicative of AGN. This arcsecond-VLBI combination favours systems aligned to the line-of-sight, thus providing a one-sided jet view. This is expected if the jets were moving at relativistic velocities. However, any entrapment of materials in the nuclear region results in the disruption of the flow of these jets, decelerating the jet plasma which results in the appearance of two-sided jets. The intermediate resolutions offered by this survey on scales of parsecs to a few sub-kpc, offer an insight into this processes where the source plasma is slowed down in the surrounding medium resulting into double-sided jets.

Advances in the wide-field VLBI technique and its application opens up possibilities for new discoveries of important astrophysical objects at mas scales, over wide areas. An increase in the sensitivity of radio surveys and the increasing ability to probe the dynamic radio sky at VLBI $\sim$ mas scales, will directly probe the missing population of dual-AGN and binary SBMHs. The SKA-VLBI will play a crucial role in these probes. The European VLBI Network and \emerlin{} are key players in these surveys since the angular resolutions provided by the EVN+\emerlin{} array, and the recovered spatial scales, will not be matched until SKA Phase 2 (SKA-VLBI). This EVN+\emerlin{} survey of the SPARCS field provides a much needed source classification training set for the near-future deep-wide field VLBI surveys with instruments such as the MeerKAT and the SKA.

\section*{Acknowledgements}
The authors are grateful to the anonymous reviewer for their insightful and constructive feedback which improved the quality of the final manuscript.
Research reported in this publication was supported by a Newton Fund project, DARA (Development in Africa with Radio Astronomy), and awarded by the UK’s Science and Technology Facilities Council (STFC) - grant reference ST/R001103/1. The European VLBI Network is a joint facility of independent European, African, Asian, and North American radio astronomy institutes. Scientific results from data presented in this publication are derived from the following EVN project code:  EN004. The research leading to these results has received funding from the European Commission Horizon 2020 Research and Innovation Programme under grant agreement No. 730562 (RadioNet). \emerlin{} and formerly, MERLIN, is a National Facility operated by the University of Manchester at Jodrell Bank Observatory on behalf of STFC. We thank IRIS \href{www.iris.ac.uk}{www.iris.ac.uk} for provision of high-performance computing facilities. STFC IRIS is investing in the UK’s Radio and mm/sub-mm Interferometry Services to improve data quality and throughput. We acknowledge the use of the ilifu cloud computing facility - \href{www.ilifu.ac.za}{www.ilifu.ac.za}, a partnership between the University of Cape Town, the University of the Western Cape, the University of Stellenbosch, Sol Plaatje University, the Cape Peninsula University of Technology and the South African Radio Astronomy Observatory. 
Javier Moldon acknowledges financial support from the State Agency for Research of the Spanish MCIU through the ``Center of Excellence Severo Ochoa'' award to the Instituto de Astrof\'isica de Andaluc\'ia (SEV-2017-0709), from the grant RTI2018-096228-B-C31 (MICIU/FEDER, EU) and the Grant No. IAA4SKA P18-RT-3082 (Regional Government of Andalusia).


\section*{Data Availability}

Data underlying this article are publicly available in the EVN Data Archive at JIVE at \href{https://www.jive.eu/select-experiment}{www.jive.eu/select-experiment} and can be accessed with project code EN004. The reduced data will be shared on reasonable request to the corresponding author.


\bibliographystyle{mnras}
\bibliography{sparcs}

\begin{thebibliography}{}
\makeatletter
\relax
\def\mn@urlcharsother{\let\do\@makeother \do\$\do\&\do\#\do\^\do\_\do\%\do\~}
\def\mn@doi{\begingroup\mn@urlcharsother \@ifnextchar [ {\mn@doi@}
  {\mn@doi@[]}}
\def\mn@doi@[#1]#2{\def\@tempa{#1}\ifx\@tempa\@empty \href
  {http://dx.doi.org/#2} {doi:#2}\else \href {http://dx.doi.org/#2} {#1}\fi
  \endgroup}
\def\mn@eprint#1#2{\mn@eprint@#1:#2::\@nil}
\def\mn@eprint@arXiv#1{\href {http://arxiv.org/abs/#1} {{\tt arXiv:#1}}}
\def\mn@eprint@dblp#1{\href {http://dblp.uni-trier.de/rec/bibtex/#1.xml}
  {dblp:#1}}
\def\mn@eprint@#1:#2:#3:#4\@nil{\def\@tempa {#1}\def\@tempb {#2}\def\@tempc
  {#3}\ifx \@tempc \@empty \let \@tempc \@tempb \let \@tempb \@tempa \fi \ifx
  \@tempb \@empty \def\@tempb {arXiv}\fi \@ifundefined
  {mn@eprint@\@tempb}{\@tempb:\@tempc}{\expandafter \expandafter \csname
  mn@eprint@\@tempb\endcsname \expandafter{\@tempc}}}

\bibitem[\protect\citeauthoryear{{Algera} et~al.,}{{Algera}
  et~al.}{2020}]{Algera2020}
{Algera} H.~S.~B.,  et~al., 2020, \mn@doi [\apj] {10.3847/1538-4357/abb77a},
  \href {https://ui.adsabs.harvard.edu/abs/2020ApJ...903..139A} {903, 139}

\bibitem[\protect\citeauthoryear{{An} et~al.,}{{An} et~al.}{2021}]{An2021}
{An} F.,  et~al., 2021, \mn@doi [\mnras] {10.1093/mnras/stab2290}, \href
  {https://ui.adsabs.harvard.edu/abs/2021MNRAS.507.2643A} {507, 2643}

\bibitem[\protect\citeauthoryear{{Arca-Sedda} \&
  {Capuzzo-Dolcetta}}{{Arca-Sedda} \&
  {Capuzzo-Dolcetta}}{2019}]{Arca-Sedda2019}
{Arca-Sedda} M.,  {Capuzzo-Dolcetta} R.,  2019, \mn@doi [\mnras]
  {10.1093/mnras/sty3096}, \href
  {https://ui.adsabs.harvard.edu/abs/2019MNRAS.483..152A} {483, 152}

\bibitem[\protect\citeauthoryear{{Blecha}, {Loeb}  \& {Narayan}}{{Blecha}
  et~al.}{2013}]{Blecha2013}
{Blecha} L.,  {Loeb} A.,   {Narayan} R.,  2013, \mn@doi [\mnras]
  {10.1093/mnras/sts533}, \href
  {https://ui.adsabs.harvard.edu/abs/2013MNRAS.429.2594B} {429, 2594}

\bibitem[\protect\citeauthoryear{{Blundell} \& {Kuncic}}{{Blundell} \&
  {Kuncic}}{2007}]{Blundell2007}
{Blundell} K.~M.,  {Kuncic} Z.,  2007, \mn@doi [\apjl] {10.1086/522695}, \href
  {https://ui.adsabs.harvard.edu/abs/2007ApJ...668L.103B} {668, L103}

\bibitem[\protect\citeauthoryear{{Breiding}, {Burke-Spolaor}, {An}, {Bansal},
  {Mohan}, {Taylor}  \& {Zhang}}{{Breiding} et~al.}{2022}]{Breiding2022}
{Breiding} P.,  {Burke-Spolaor} S.,  {An} T.,  {Bansal} K.,  {Mohan} P.,
  {Taylor} G.~B.,   {Zhang} Y.,  2022, \mn@doi [\apj]
  {10.3847/1538-4357/ac7466}, \href
  {https://ui.adsabs.harvard.edu/abs/2022ApJ...933..143B} {933, 143}

\bibitem[\protect\citeauthoryear{{Burke-Spolaor}}{{Burke-Spolaor}}{2011}]{Burke2011}
{Burke-Spolaor} S.,  2011, \mn@doi [\mnras] {10.1111/j.1365-2966.2010.17586.x},
  \href {https://ui.adsabs.harvard.edu/abs/2011MNRAS.410.2113B} {410, 2113}

\bibitem[\protect\citeauthoryear{{Burke-Spolaor}, {Brazier}, {Chatterjee},
  {Comerford}, {Cordes}, {Lazio}, {Liu}  \& {Shen}}{{Burke-Spolaor}
  et~al.}{2014}]{Burke2014}
{Burke-Spolaor} S.,  {Brazier} A.,  {Chatterjee} S.,  {Comerford} J.,  {Cordes}
  J.,  {Lazio} T. J.~W.,  {Liu} X.,   {Shen} Y.,  2014, arXiv e-prints, \href
  {https://ui.adsabs.harvard.edu/abs/2014arXiv1402.0548B} {p. arXiv:1402.0548}

\bibitem[\protect\citeauthoryear{{Burke-Spolaor} et~al.,}{{Burke-Spolaor}
  et~al.}{2018}]{Burke2018}
{Burke-Spolaor} S.,  et~al., 2018, arXiv e-prints, \href
  {https://ui.adsabs.harvard.edu/abs/2018arXiv180804368B} {p. arXiv:1808.04368}

\bibitem[\protect\citeauthoryear{{Chi}, {Barthel}  \& {Garrett}}{{Chi}
  et~al.}{2013}]{Chi2013}
{Chi} S.,  {Barthel} P.~D.,   {Garrett} M.~A.,  2013, \mn@doi [\aap]
  {10.1051/0004-6361/201220783}, \href
  {https://ui.adsabs.harvard.edu/abs/2013A&A...550A..68C} {550, A68}

\bibitem[\protect\citeauthoryear{{Condon}, {Huang}, {Yin}  \& {Thuan}}{{Condon}
  et~al.}{1991}]{Condon1991}
{Condon} J.~J.,  {Huang} Z.~P.,  {Yin} Q.~F.,   {Thuan} T.~X.,  1991, \mn@doi
  [\apj] {10.1086/170407}, \href
  {https://ui.adsabs.harvard.edu/abs/1991ApJ...378...65C} {378, 65}

\bibitem[\protect\citeauthoryear{{Cotton} \& {Mauch}}{{Cotton} \&
  {Mauch}}{2021}]{CottonMauch2021}
{Cotton} W.~D.,  {Mauch} T.,  2021, \mn@doi [\pasp] {10.1088/1538-3873/ac2351},
  \href {https://ui.adsabs.harvard.edu/abs/2021PASP..133j4502C} {133, 104502}

\bibitem[\protect\citeauthoryear{{Croston} et~al.,}{{Croston}
  et~al.}{2019}]{Croston2019}
{Croston} J.~H.,  et~al., 2019, \mn@doi [\aap] {10.1051/0004-6361/201834019},
  \href {https://ui.adsabs.harvard.edu/abs/2019A&A...622A..10C} {622, A10}

\bibitem[\protect\citeauthoryear{{De Rosa} et~al.,}{{De Rosa}
  et~al.}{2019}]{DeRosa2019}
{De Rosa} A.,  et~al., 2019, \mn@doi [\nar] {10.1016/j.newar.2020.101525},
  \href {https://ui.adsabs.harvard.edu/abs/2019NewAR..8601525D} {86, 101525}

\bibitem[\protect\citeauthoryear{{Deane} et~al.,}{{Deane}
  et~al.}{2014}]{Deane2014}
{Deane} R.~P.,  et~al., 2014, \mn@doi [\nat] {10.1038/nature13454}, \href
  {https://ui.adsabs.harvard.edu/abs/2014Natur.511...57D} {511, 57}

\bibitem[\protect\citeauthoryear{{Deller} \& {Middelberg}}{{Deller} \&
  {Middelberg}}{2014}]{Deller2014}
{Deller} A.~T.,  {Middelberg} E.,  2014, \mn@doi [\aj]
  {10.1088/0004-6256/147/1/14}, \href
  {https://ui.adsabs.harvard.edu/abs/2014AJ....147...14D} {147, 14}

\bibitem[\protect\citeauthoryear{{Hales}, {Murphy}, {Curran}, {Middelberg},
  {Gaensler}  \& {Norris}}{{Hales} et~al.}{2012}]{Hales2012}
{Hales} C.~A.,  {Murphy} T.,  {Curran} J.~R.,  {Middelberg} E.,  {Gaensler}
  B.~M.,   {Norris} R.~P.,  2012, \mn@doi [\mnras]
  {10.1111/j.1365-2966.2012.21373.x}, \href
  {https://ui.adsabs.harvard.edu/abs/2012MNRAS.425..979H} {425, 979}

\bibitem[\protect\citeauthoryear{{Hardcastle} \& {Croston}}{{Hardcastle} \&
  {Croston}}{2020}]{Hardcastle2020}
{Hardcastle} M.~J.,  {Croston} J.~H.,  2020, \mn@doi [\nar]
  {10.1016/j.newar.2020.101539}, \href
  {https://ui.adsabs.harvard.edu/abs/2020NewAR..8801539H} {88, 101539}

\bibitem[\protect\citeauthoryear{{Hardcastle} et~al.,}{{Hardcastle}
  et~al.}{2019}]{Hardcastle2019}
{Hardcastle} M.~J.,  et~al., 2019, \mn@doi [\aap]
  {10.1051/0004-6361/201833893}, \href
  {https://ui.adsabs.harvard.edu/abs/2019A&A...622A..12H} {622, A12}

\bibitem[\protect\citeauthoryear{{Herrera-Ruiz} et~al.,}{{Herrera-Ruiz}
  et~al.}{2017}]{Herrera2017}
{Herrera-Ruiz} N.,  et~al., 2017, \mn@doi [aap] {10.1051/0004-6361/201731163},
  \href {https://ui.adsabs.harvard.edu/abs/2017A&A...607A.132H} {607, A132}

\bibitem[\protect\citeauthoryear{{Herrera-Ruiz} et~al.,}{{Herrera-Ruiz}
  et~al.}{2018}]{Herrera2018}
{Herrera-Ruiz} N.,  et~al., 2018, \mn@doi [\aap] {10.1051/0004-6361/201832969},
  \href {https://ui.adsabs.harvard.edu/abs/2018A&A...616A.128H} {616, A128}

\bibitem[\protect\citeauthoryear{{Keimpema} et~al.,}{{Keimpema}
  et~al.}{2015}]{Keimpema2015}
{Keimpema} A.,  et~al., 2015, \mn@doi [Experimental Astronomy]
  {10.1007/s10686-015-9446-1}, \href
  {https://ui.adsabs.harvard.edu/abs/2015ExA....39..259K} {39, 259}

\bibitem[\protect\citeauthoryear{{Kellermann}}{{Kellermann}}{1963}]{Kellermann1963}
{Kellermann} K.~I.,  1963, \mn@doi [\aj] {10.1086/109010}, \href
  {https://ui.adsabs.harvard.edu/abs/1963AJ.....68..539K} {68, 539}

\bibitem[\protect\citeauthoryear{{Kondapally} et~al.,}{{Kondapally}
  et~al.}{2021}]{Kondapally2021}
{Kondapally} R.,  et~al., 2021, \mn@doi [\aap] {10.1051/0004-6361/202038813},
  \href {https://ui.adsabs.harvard.edu/abs/2021A&A...648A...3K} {648, A3}

\bibitem[\protect\citeauthoryear{{Lacy} et~al.,}{{Lacy}
  et~al.}{2020}]{Lacy2020}
{Lacy} M.,  et~al., 2020, \mn@doi [\pasp] {10.1088/1538-3873/ab63eb}, \href
  {https://ui.adsabs.harvard.edu/abs/2020PASP..132c5001L} {132, 035001}

\bibitem[\protect\citeauthoryear{{Law} et~al.,}{{Law} et~al.}{2021}]{Law2021}
{Law} D.~R.,  et~al., 2021, \mn@doi [\apj] {10.3847/1538-4357/abfe0a}, \href
  {https://ui.adsabs.harvard.edu/abs/2021ApJ...915...35L} {915, 35}

\bibitem[\protect\citeauthoryear{{Macfarlane} et~al.,}{{Macfarlane}
  et~al.}{2021}]{Macfarlane2021}
{Macfarlane} C.,  et~al., 2021, \mn@doi [\mnras] {10.1093/mnras/stab1998},
  \href {https://ui.adsabs.harvard.edu/abs/2021MNRAS.506.5888M} {506, 5888}

\bibitem[\protect\citeauthoryear{{Middelberg} \& {Bach}}{{Middelberg} \&
  {Bach}}{2008}]{Middelberg2008}
{Middelberg} E.,  {Bach} U.,  2008, \mn@doi [Reports on Progress in Physics]
  {10.1088/0034-4885/71/6/066901}, \href
  {https://ui.adsabs.harvard.edu/abs/2008RPPh...71f6901M} {71, 066901}

\bibitem[\protect\citeauthoryear{{Middelberg} et~al.,}{{Middelberg}
  et~al.}{2013}]{Middelberg2013}
{Middelberg} E.,  et~al., 2013, \mn@doi [\aap] {10.1051/0004-6361/201220374},
  \href {https://ui.adsabs.harvard.edu/abs/2013A&A...551A..97M} {551, A97}

\bibitem[\protect\citeauthoryear{{Mingarelli}}{{Mingarelli}}{2019}]{Mingarelli2019}
{Mingarelli} C. M.~F.,  2019, \mn@doi [Nature Astronomy]
  {10.1038/s41550-018-0666-y}, \href
  {https://ui.adsabs.harvard.edu/abs/2019NatAs...3....8M} {3, 8}

\bibitem[\protect\citeauthoryear{{Mingo} et~al.,}{{Mingo}
  et~al.}{2019}]{Mingo2019}
{Mingo} B.,  et~al., 2019, \mn@doi [\mnras] {10.1093/mnras/stz1901}, \href
  {https://ui.adsabs.harvard.edu/abs/2019MNRAS.488.2701M} {488, 2701}

\bibitem[\protect\citeauthoryear{{Mohan} \& {Rafferty}}{{Mohan} \&
  {Rafferty}}{2015}]{Mohan2015}
{Mohan} N.,  {Rafferty} D.,  2015, {PyBDSF: Python Blob Detection and Source
  Finder} (\mn@eprint {ascl} {1502.007})

\bibitem[\protect\citeauthoryear{{Morabito} et~al.,}{{Morabito}
  et~al.}{2017}]{Morabito2017}
{Morabito} L.~K.,  et~al., 2017, \mn@doi [\mnras] {10.1093/mnras/stx959}, \href
  {https://ui.adsabs.harvard.edu/abs/2017MNRAS.469.1883M} {469, 1883}

\bibitem[\protect\citeauthoryear{{Morabito} et~al.,}{{Morabito}
  et~al.}{2022}]{Morabito2022}
{Morabito} L.~K.,  et~al., 2022, \mn@doi [\mnras] {10.1093/mnras/stac2129},
  \href {https://ui.adsabs.harvard.edu/abs/2022MNRAS.515.5758M} {515, 5758}

\bibitem[\protect\citeauthoryear{{Muxlow} et~al.,}{{Muxlow}
  et~al.}{2020}]{Muxlow2020}
{Muxlow} T.~W.~B.,  et~al., 2020, \mn@doi [\mnras] {10.1093/mnras/staa1279},
  \href {https://ui.adsabs.harvard.edu/abs/2020MNRAS.495.1188M} {495, 1188}

\bibitem[\protect\citeauthoryear{{Norris}}{{Norris}}{2017}]{Norris2017}
{Norris} R.~P.,  2017, \mn@doi [Nature Astronomy] {10.1038/s41550-017-0233-y},
  \href {https://ui.adsabs.harvard.edu/abs/2017NatAs...1..671N} {1, 671}

\bibitem[\protect\citeauthoryear{{Norris} et~al.,}{{Norris}
  et~al.}{2013}]{Norris2013}
{Norris} R.~P.,  et~al., 2013, \mn@doi [\pasa] {10.1017/pas.2012.020}, \href
  {https://ui.adsabs.harvard.edu/abs/2013PASA...30...20N} {30, e020}

\bibitem[\protect\citeauthoryear{{Nyland} et~al.,}{{Nyland}
  et~al.}{2020}]{Nyland2020}
{Nyland} K.,  et~al., 2020, \mn@doi [\apj] {10.3847/1538-4357/abc341}, \href
  {https://ui.adsabs.harvard.edu/abs/2020ApJ...905...74N} {905, 74}

\bibitem[\protect\citeauthoryear{{Offringa}, {de Bruyn}  \&
  {Zaroubi}}{{Offringa} et~al.}{2012}]{Offringa2012}
{Offringa} A.~R.,  {de Bruyn} A.~G.,   {Zaroubi} S.,  2012, \mn@doi [\mnras]
  {10.1111/j.1365-2966.2012.20633.x}, \href
  {https://ui.adsabs.harvard.edu/abs/2012MNRAS.422..563O} {422, 563}

\bibitem[\protect\citeauthoryear{{Padovani}}{{Padovani}}{2016}]{Padovani2016}
{Padovani} P.,  2016, \mn@doi [\aapr] {10.1007/s00159-016-0098-6}, \href
  {http://adsabs.harvard.edu/abs/2016A%26ARv..24...13P} {24, 13}

\bibitem[\protect\citeauthoryear{{Planck Collaboration} et~al.,}{{Planck
  Collaboration} et~al.}{2020}]{Planck2020a}
{Planck Collaboration} et~al., 2020, \mn@doi [\aap]
  {10.1051/0004-6361/201833910}, \href
  {https://ui.adsabs.harvard.edu/abs/2020A&A...641A...6P} {641, A6}

\bibitem[\protect\citeauthoryear{{Radcliffe}, {Garrett}, {Beswick}, {Muxlow},
  {Barthel}, {Deller}  \& {Middelberg}}{{Radcliffe}
  et~al.}{2016}]{Radcliffe2016}
{Radcliffe} J.~F.,  {Garrett} M.~A.,  {Beswick} R.~J.,  {Muxlow} T.~W.~B.,
  {Barthel} P.~D.,  {Deller} A.~T.,   {Middelberg} E.,  2016, \mn@doi [\aap]
  {10.1051/0004-6361/201527980}, \href
  {https://ui.adsabs.harvard.edu/abs/2016A&A...587A..85R} {587, A85}

\bibitem[\protect\citeauthoryear{{Radcliffe} et~al.,}{{Radcliffe}
  et~al.}{2018}]{Radcliffe2018}
{Radcliffe} J.~F.,  et~al., 2018, \mn@doi [\aap] {10.1051/0004-6361/201833399},
  \href {https://ui.adsabs.harvard.edu/abs/2018AA...619A..48R} {619, A48}

\bibitem[\protect\citeauthoryear{{Radcliffe} et~al.,}{{Radcliffe}
  et~al.}{2019}]{Radcliffe2019}
{Radcliffe} J.~F.,  et~al., 2019, \mn@doi [\aap]
  {10.1051/0004-6361/201833399e}, \href
  {https://ui.adsabs.harvard.edu/abs/2019A&A...625C...1R} {625, C1}

\bibitem[\protect\citeauthoryear{{Radcliffe}, {Barthel}, {Thomson}, {Garrett},
  {Beswick}  \& {Muxlow}}{{Radcliffe} et~al.}{2021}]{Radcliffe2021}
{Radcliffe} J.~F.,  {Barthel} P.~D.,  {Thomson} A.~P.,  {Garrett} M.~A.,
  {Beswick} R.~J.,   {Muxlow} T.~W.~B.,  2021, \mn@doi [\aap]
  {10.1051/0004-6361/202038591}, \href
  {https://ui.adsabs.harvard.edu/abs/2021A&A...649A..27R} {649, A27}

\bibitem[\protect\citeauthoryear{{Rawlings} et~al.,}{{Rawlings}
  et~al.}{2015}]{Rawlings2015}
{Rawlings} J.~I.,  et~al., 2015, \mn@doi [\mnras] {10.1093/mnras/stv1573},
  \href {https://ui.adsabs.harvard.edu/abs/2015MNRAS.452.4111R} {452, 4111}

\bibitem[\protect\citeauthoryear{{Rees}, {Norris}, {Spitler}, {Herrera-Ruiz}
  \& {Middelberg}}{{Rees} et~al.}{2016}]{Rees2016}
{Rees} G.~A.,  {Norris} R.~P.,  {Spitler} L.~R.,  {Herrera-Ruiz} N.,
  {Middelberg} E.,  2016, \mn@doi [\mnras] {10.1093/mnrasl/slw016}, \href
  {https://ui.adsabs.harvard.edu/abs/2016MNRAS.458L..49R} {458, L49}

\bibitem[\protect\citeauthoryear{{Sarbadhicary} et~al.,}{{Sarbadhicary}
  et~al.}{2021}]{Sarbadhicary2021}
{Sarbadhicary} S.~K.,  et~al., 2021, \mn@doi [\apj] {10.3847/1538-4357/ac2239},
  \href {https://ui.adsabs.harvard.edu/abs/2021ApJ...923...31S} {923, 31}

\bibitem[\protect\citeauthoryear{{Simpson}}{{Simpson}}{2017}]{Simpson2017}
{Simpson} C.,  2017, \mn@doi [Royal Society Open Science]
  {10.1098/rsos.170522}, \href
  {https://ui.adsabs.harvard.edu/abs/2017RSOS....470522S} {4, 170522}

\bibitem[\protect\citeauthoryear{{Slijepcevic}, {Scaife}, {Walmsley}, {Bowles},
  {Wong}, {Shabala}  \& {Tang}}{{Slijepcevic} et~al.}{2022}]{Inigo2022}
{Slijepcevic} I.~V.,  {Scaife} A. M.~M.,  {Walmsley} M.,  {Bowles} M.,  {Wong}
  O.~I.,  {Shabala} S.~S.,   {Tang} H.,  2022, \mn@doi [\mnras]
  {10.1093/mnras/stac1135}, \href
  {https://ui.adsabs.harvard.edu/abs/2022MNRAS.514.2599S} {514, 2599}

\bibitem[\protect\citeauthoryear{{Spingola}, {McKean}, {Auger}, {Fassnacht},
  {Koopmans}, {Lagattuta}  \& {Vegetti}}{{Spingola}
  et~al.}{2018}]{Spingola2018}
{Spingola} C.,  {McKean} J.~P.,  {Auger} M.~W.,  {Fassnacht} C.~D.,  {Koopmans}
  L.~V.~E.,  {Lagattuta} D.~J.,   {Vegetti} S.,  2018, \mn@doi [\mnras]
  {10.1093/mnras/sty1326}, \href
  {https://ui.adsabs.harvard.edu/abs/2018MNRAS.478.4816S} {478, 4816}

\bibitem[\protect\citeauthoryear{{Spingola}, {McKean}, {Lee}, {Deller}  \&
  {Moldon}}{{Spingola} et~al.}{2019}]{Spingola2019}
{Spingola} C.,  {McKean} J.~P.,  {Lee} M.,  {Deller} A.,   {Moldon} J.,  2019,
  \mn@doi [\mnras] {10.1093/mnras/sty3189}, \href
  {https://ui.adsabs.harvard.edu/abs/2019MNRAS.483.2125S} {483, 2125}

\bibitem[\protect\citeauthoryear{{Thomson} et~al.,}{{Thomson}
  et~al.}{2014}]{Thomson2014}
{Thomson} A.~P.,  et~al., 2014, \mn@doi [\mnras] {10.1093/mnras/stu839}, \href
  {https://ui.adsabs.harvard.edu/abs/2014MNRAS.442..577T} {442, 577}

\bibitem[\protect\citeauthoryear{{Van Wassenhove}, {Volonteri}, {Mayer},
  {Dotti}, {Bellovary}  \& {Callegari}}{{Van Wassenhove}
  et~al.}{2012}]{Wassenhove2012}
{Van Wassenhove} S.,  {Volonteri} M.,  {Mayer} L.,  {Dotti} M.,  {Bellovary}
  J.,   {Callegari} S.,  2012, \mn@doi [\apjl] {10.1088/2041-8205/748/1/L7},
  \href {https://ui.adsabs.harvard.edu/abs/2012ApJ...748L...7V} {748, L7}

\bibitem[\protect\citeauthoryear{{Vernstrom}, {Scott}, {Wall}, {Condon},
  {Cotton}  \& {Perley}}{{Vernstrom} et~al.}{2016}]{Vernstrom2016}
{Vernstrom} T.,  {Scott} D.,  {Wall} J.~V.,  {Condon} J.~J.,  {Cotton} W.~D.,
  {Perley} R.~A.,  2016, \mn@doi [\mnras] {10.1093/mnras/stw1530}, \href
  {http://adsabs.harvard.edu/abs/2016MNRAS.461.2879V} {461, 2879}

\bibitem[\protect\citeauthoryear{{Whittam} et~al.,}{{Whittam}
  et~al.}{2022}]{Whittam2022}
{Whittam} I.~H.,  et~al., 2022, \mn@doi [\mnras] {10.1093/mnras/stac2140},
  \href {https://ui.adsabs.harvard.edu/abs/2022MNRAS.516..245W} {516, 245}

\bibitem[\protect\citeauthoryear{{van der Vlugt} et~al.,}{{van der Vlugt}
  et~al.}{2021}]{Vlugt2021}
{van der Vlugt} D.,  et~al., 2021, \mn@doi [\apj] {10.3847/1538-4357/abcaa3},
  \href {https://ui.adsabs.harvard.edu/abs/2021ApJ...907....5V} {907, 5}

\makeatother
\end{thebibliography}




\bsp	
\label{lastpage}
\end{document}